\documentclass{caosp308}
\usepackage{graphicx}
\usepackage{natbib}
\usepackage{booktabs}
 \usepackage{url}
\usepackage[table]{xcolor}
\usepackage{array}
\usepackage{float}
\usepackage{multirow}
\usepackage{multicol}
\usepackage{blindtext}
\usepackage{longtable}
\usepackage{subcaption}
\definecolor{cyan}{rgb}{0.88,1.0,1.0}
\definecolor{yellow}{rgb}{1.0,1.0,0.0}
\newcommand{\Msun}{\ensuremath{\rm M_\odot}}

\articleNo{123}
\pubyear{2020}
\volume{XX}
\volnumber{Y}
\firstpage{Z}
\received{September ??, 2020}
\accepted{????????? ??, 2020}

\def\BibTeX{{\rm B\kern-.05em{\sc i\kern-.025em b}\kern-.08em
             T\kern-.1667em\lower.7ex\hbox{E}\kern-.125emX}}

\begin{document}

\title{Flare stars in nearby Galactic open clusters based on {\it TESS} data}
\hauthor{O.\,Maryeva et al.}
\author{Olga\,Maryeva\inst{1,2}\orcid{0000-0003-1442-4755}
      \and
        Kamil\,Bicz\inst{3} 
      \and 
        Caiyun\,Xia\inst{4}
      \and
        Martina\,Baratella\inst{5}\orcid{https://orcid.org/0000-0002-1027-5003}
      \and 
        Patrik\,Čechvala\inst{6}   
      \and 
        Krisztian\,Vida\inst{7}
       }
%
\institute{
            \ondrejov \email{olga.maryeva@asu.cas.cz}
         \and
            Lomonosov Moscow State University, Sternberg Astronomical Institute,\\ 
            Universitetsky pr. 13, 119234, Moscow, Russia 
         \and
         Astronomical Institute, University of Wroc\l aw, Kopernika 11, 51-622 Wroc\l aw, Poland 
         \and
            Department of Theoretical Physics and Astrophysics,
            Faculty of Science,\\
            Masaryk University, Kotlářská 2, 611 37 Brno, Czech Republic
         \and 
           Dipartimento di Fisica e Astronomia Galileo Galilei, Vicolo Osservatorio 3, 35122, Padova, Italy, \email{martina.baratella.1@phd.unipd.it}
         \and 
           Department of Astronomy, Physics of the Earth and Meteorology, Faculty of Mathematics, Physics and Informatics,\\
           Comenius University in Bratislava, Mlynská dolina F-2, 842 48 Bratislava, Slovakia
         \and 
         Konkoly Observatory, Research Centre for Astronomy and Earth Sciences,\\ H-1121 Budapest, Konkoly Thege Mikl\'os \'ut 15-17, Hungary\\
          }
          
\date{September 2020}
\maketitle
\begin{abstract} 
The study is devoted to search for flare stars among confirmed members of Galactic open clusters using high-cadence photometry from {\it TESS} mission.
We analyzed 957 high-cadence light curves of members from 136 open clusters. 
As a result, 56 flare stars were found, among them 8 hot B-A type objects. Of all flares, 63\,\% were detected in sample of cool stars ($T_{\rm eff}<5000$\,K), and 29\,\% -- in stars of spectral type G, while 23\,\% in K-type stars and approximately 34\% of all detected flares are  in M-type stars. 
Using the \texttt{FLATW'RM} (FLAre deTection With Ransac Method) flare finding algorithm, we estimated parameters of flares and rotation period of detected flare stars. 
The flare with the largest amplitude appears on the M3 type EQ\,Cha star. 
Statistical analysis did not reveal any direct correlation between ages, rotation periods and flaring activity.  

\end{abstract}

\newpage
\section{Introduction}

Flare stars --  stars having detected at least one very short flare in their light curve -- are a widely known type of objects in astrophysics. 
The physical mechanisms leading to the appearance of flares are associated with convective atmospheres, such as in G--M type stars on the main sequence. Thus most of the stars in our Galaxy are potentially flare stars. 

In the middle of 20th century, only six flare stars were known \citep{lippincott1952}. Since then, a huge progress was made in the field of detecting stellar flares. The interest of astronomers to these objects is associated not only with the incompletely studied mechanisms of formation of flares, but also with the possible influence of flares on habitable zone of potential exoplanets. The {\it Kepler} mission had a significant impact on the study of flare stars. As a result, 
\newline
{\it The Kepler Catalog of Stellar Flares}  
with more than 4000 detected flare stars was made by \citet{davenport2016}.
And even more significant impact can be expected from the ongoing {\it Transiting Exoplanet Survey Satellite (TESS)} mission, which already covered full celestial sphere.
{\it TESS} satellite is a space-borne telescope, whose main objective is the detection and study of exoplanets using the technique of transits. However, its database of a high-cadence (2 minute effective sampling) light curves for hundreds of thousands of objects, as well as 
sequences of Full Frame Images (30 minutes effective sampling), may also provide an invaluable resource for the study of many other kinds of astrophysical objects, including flare stars.

Stellar flares are explosive magnetic reconnection events in a star’s magnetosphere \citep{gunther2020} emitting energy through radio wavelength, X-ray, UV, optical and IR band \citep{lawson2019}. They show a sharp rise in intensity followed by an exponential decay. Duration of these flares is typically within few minutes to hours \citep{doyle2018}. The presence of flares indicates magnetic activity in the stellar magnetosphere. Since the magnetic activity is strongly related to the strength of the magnetic field, fast-rotating young stars are typically more active. 
The magnetic activity also changes over time, so one would expect the frequency and power of flares to change as well. However, it is hard to investigate these changes in G-M type stars due to their long lifetime and slow evolution on the main sequence.

There are several possible ways to estimate the age of such late type stars, such as method based on the rotation period (gyrochronology, \citealt{barnes2007}), or spatial motion (kinematic group membership). But probably the most effective and affordable one is by means of estimating an age of a parent cluster. Therefore, the aim of this study is to evaluate the effect of stellar age on the activity of flare stars in Galactic open clusters. 

Open clusters are groups of stars loosely held together by their mutual gravitational attraction and formed from the same interstellar molecular cloud. Therefore the age, distance and chemical properties of members in open clusters are approximately identical. This means that the chemical composition of open clusters also reflects a real composition of individual members \citep{netopil2015}. 

In this paper, we combine the high temporal resolution light curves acquired by {\it TESS} satellite with the cluster membership data based on {\it Gaia} DR2 \citep{2018cantat}. This allows us to reliably select the members of Galactic open clusters with known age and distance, examine their variability and extract the ones with prominent stellar flares for further statistical analysis.

The paper is organized as follows. In Section~\ref{sec_2}, we briefly describe a metho-dology of selection of cluster members having {\it TESS} light curves available and the methodology of flare detection and characterisation. 
In Section~\ref{sec_3}, we present statistical analysis of detected flares. 
A conclusion of this study is given in Section~\ref{sec_4}. 

\section{Selection of flare stars}\label{sec_2}

The first step of our work was the selection of stars belonging to open clusters and being {\it TESS} targets with high temporal resolution light curves available. To do it, we took a list of Galactic open clusters from \citet{2018cantat} and selected the ones closer than 2~kpc from the Sun. Such distance cut-off allows us to discard the clusters where G0V and later type of dwarfs, the most interesting for 
flare stars statistics, are below {\it TESS} detection limit due to their intrinsic faintness. 
Then, we used the list of members from \citet{2018cantat} for all selected clusters and checked the availability of {\it TESS} light curves for every object. As a result, we acquired 859 light curves for stars located in 136 open clusters. We also considered nearby cluster Melote\,111 (or Coma Star cluster) with 98 light curves available. Basic parameters of these clusters (age, number of members, distance) are summarized in Table~\ref{OCs_list}, along with the numbers of available light curves and detected flares for them. 

As the next step, we performed a visual inspection of all these light curves and discarded the ones without any flares, thus significantly reduced the number of light curves. Then we performed an automated detection and characterisation of the flares using \texttt{FLATW'RM} (FLAre deTection With Ransac Method) code \citep{vida2018}. This code uses a machine-learning algorithm to give a robust model of the light curves in order to detect flare events and uses a voting system implemented to keep false positive detections to a minimum. \texttt{FLATW'RM} detects flares and reports the times the flare starts and ends, the time of maximum flux and the maximum percent increase of flux over the light curve around the flare, along with estimated flare energy, either from the raw light curve or by fitting an analytic flare model. It also gives a crude estimation of the period of the underlying light curve, assuming that it is periodic and using the position of the strongest peak in the Lomb-Scargle periodogram.

\begin{figure}
\centerline{\includegraphics[width=0.5\textwidth]{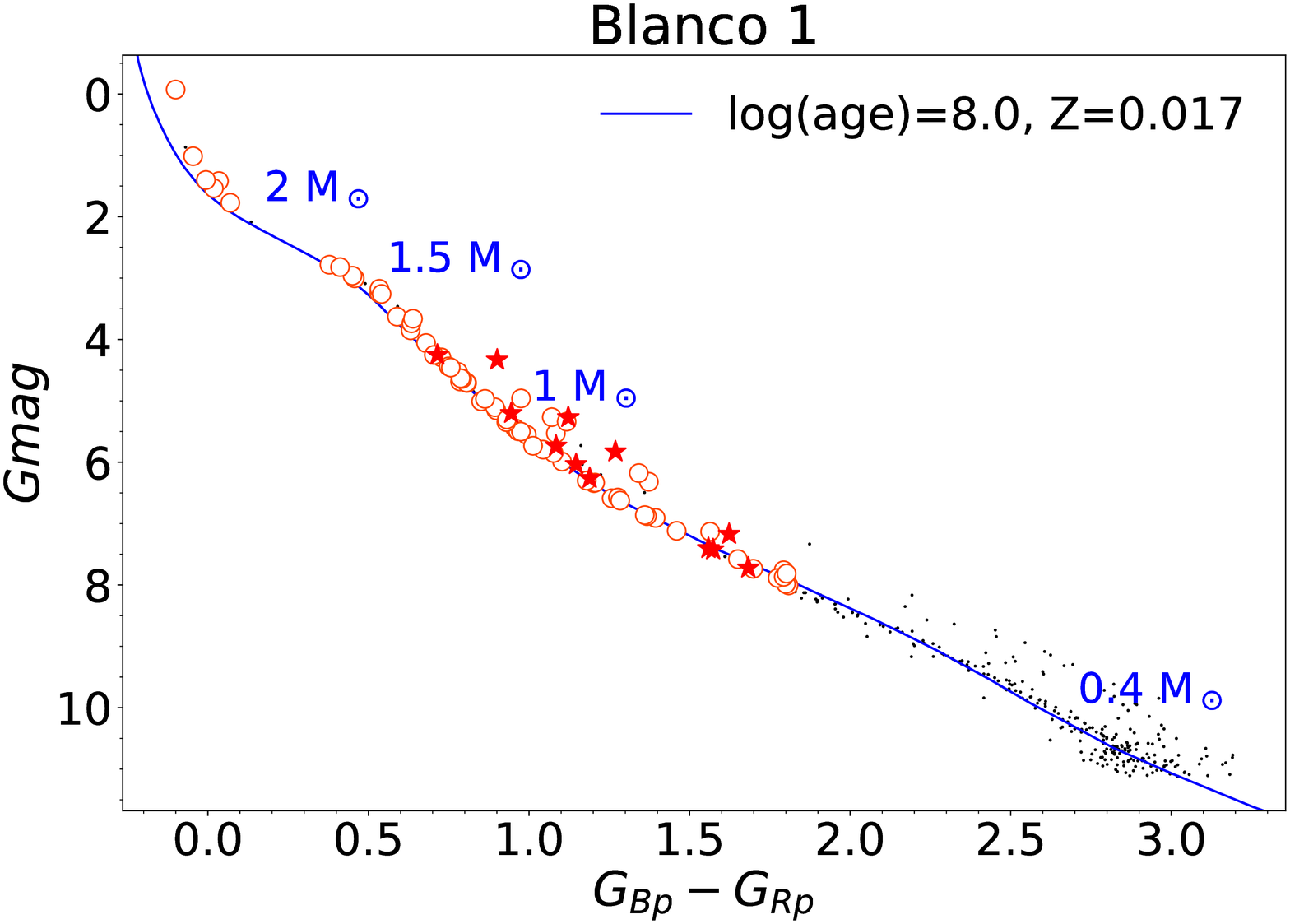}
\includegraphics[width=0.5\textwidth]{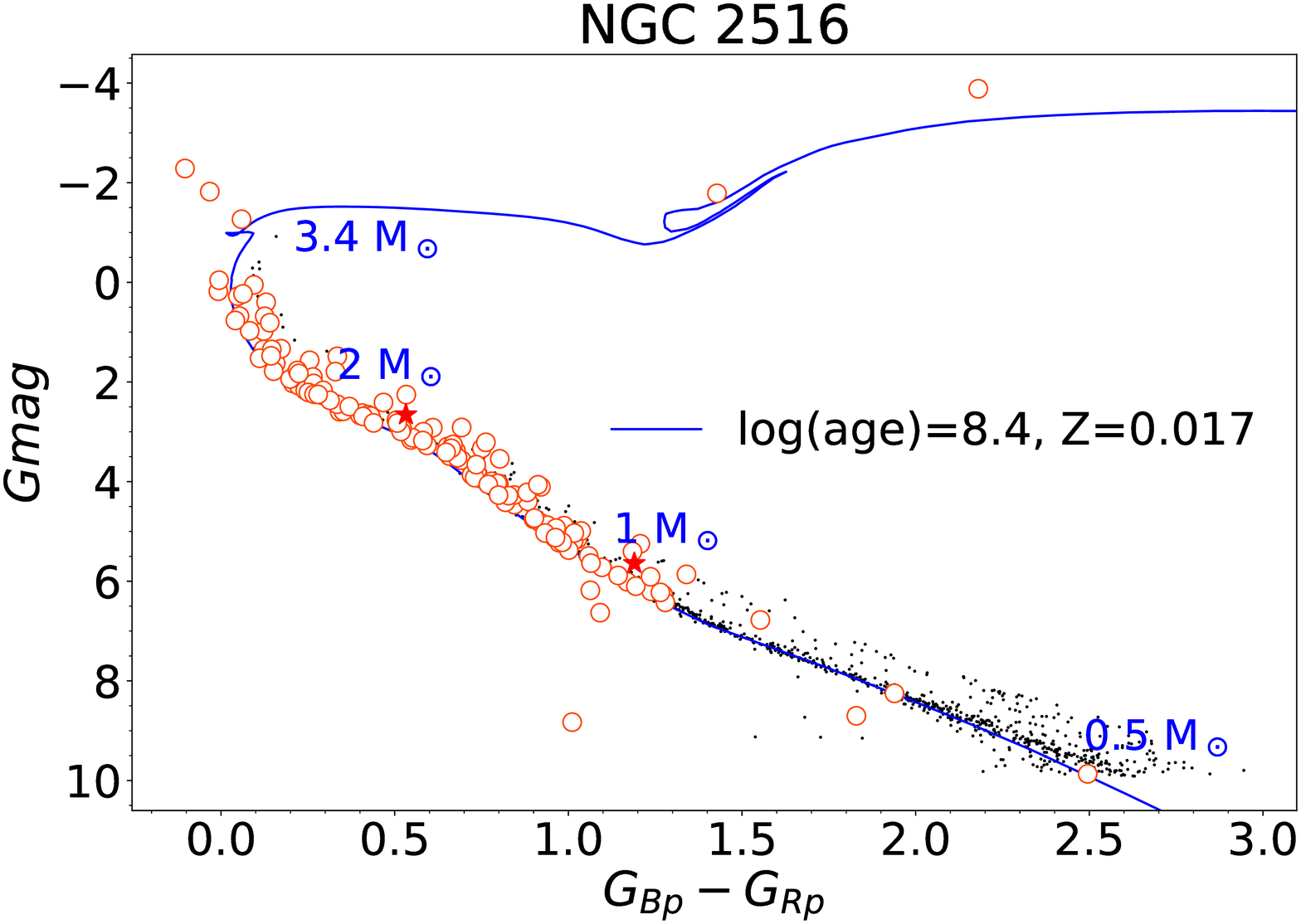}}
\centerline{\includegraphics[width=0.505\textwidth]{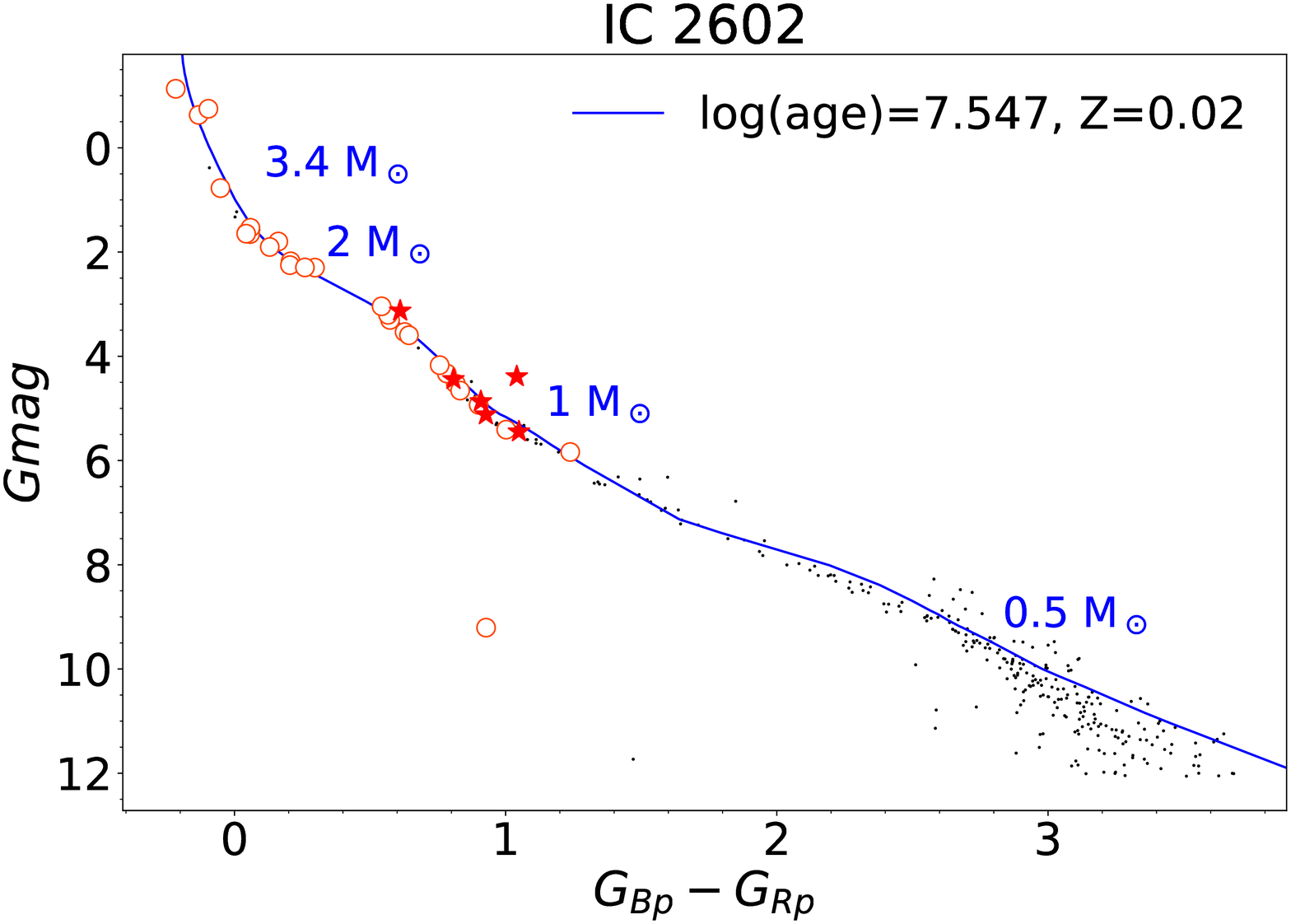}
\includegraphics[width=0.495\textwidth]{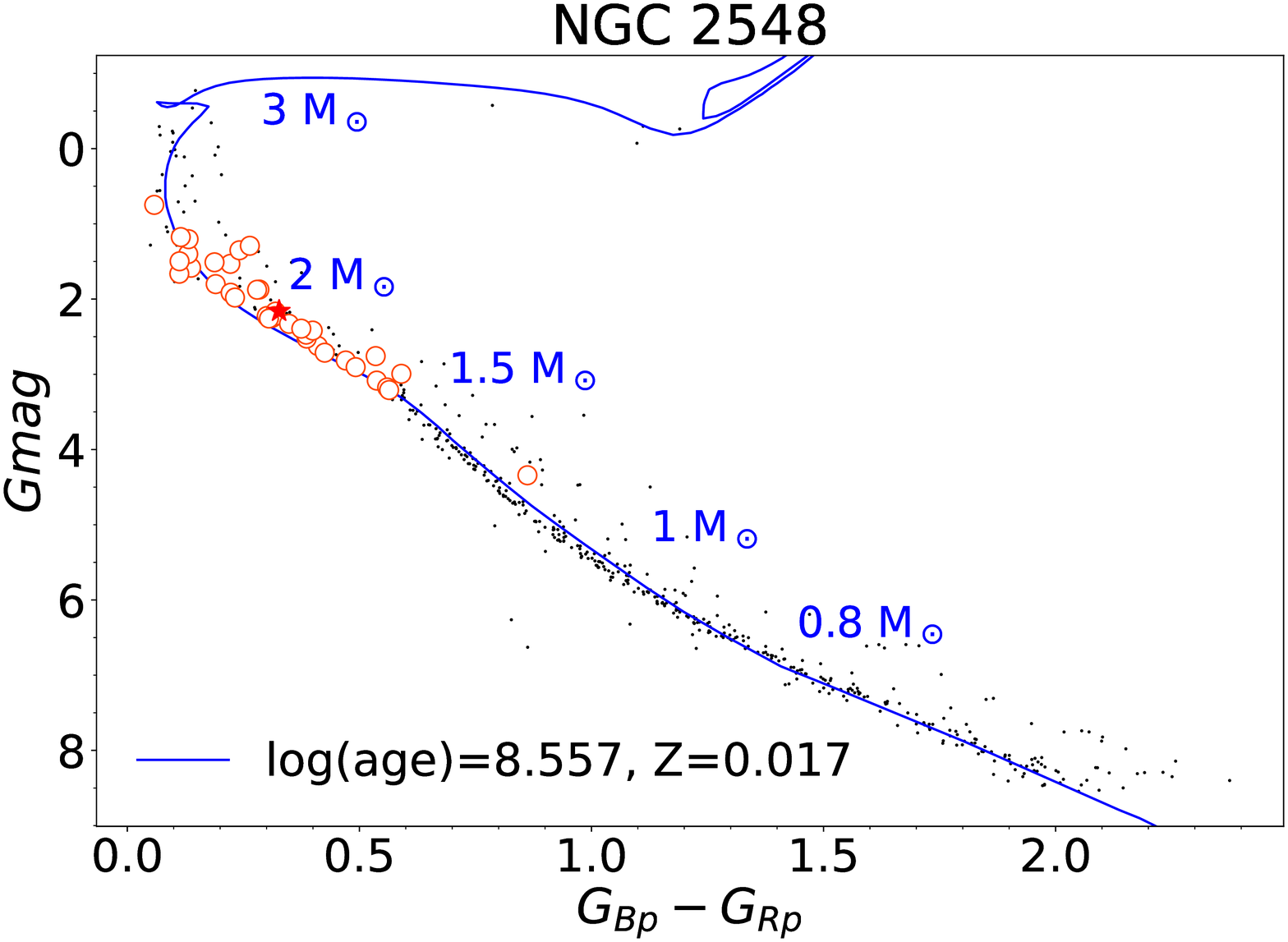}}
\caption{H--R diagram for Blanco\,1, NGC\,2516, IC\,2602 and NGC\,2548 clusters based on {\it Gaia} photometric data and compared with the PARSEC isochrones (see text for details). Red circles are TESS objects without flares, while red stars -- with flares.}
\label{hr_flash}
\end{figure}

\begin{figure}
\centerline{\includegraphics[width=0.5\textwidth]{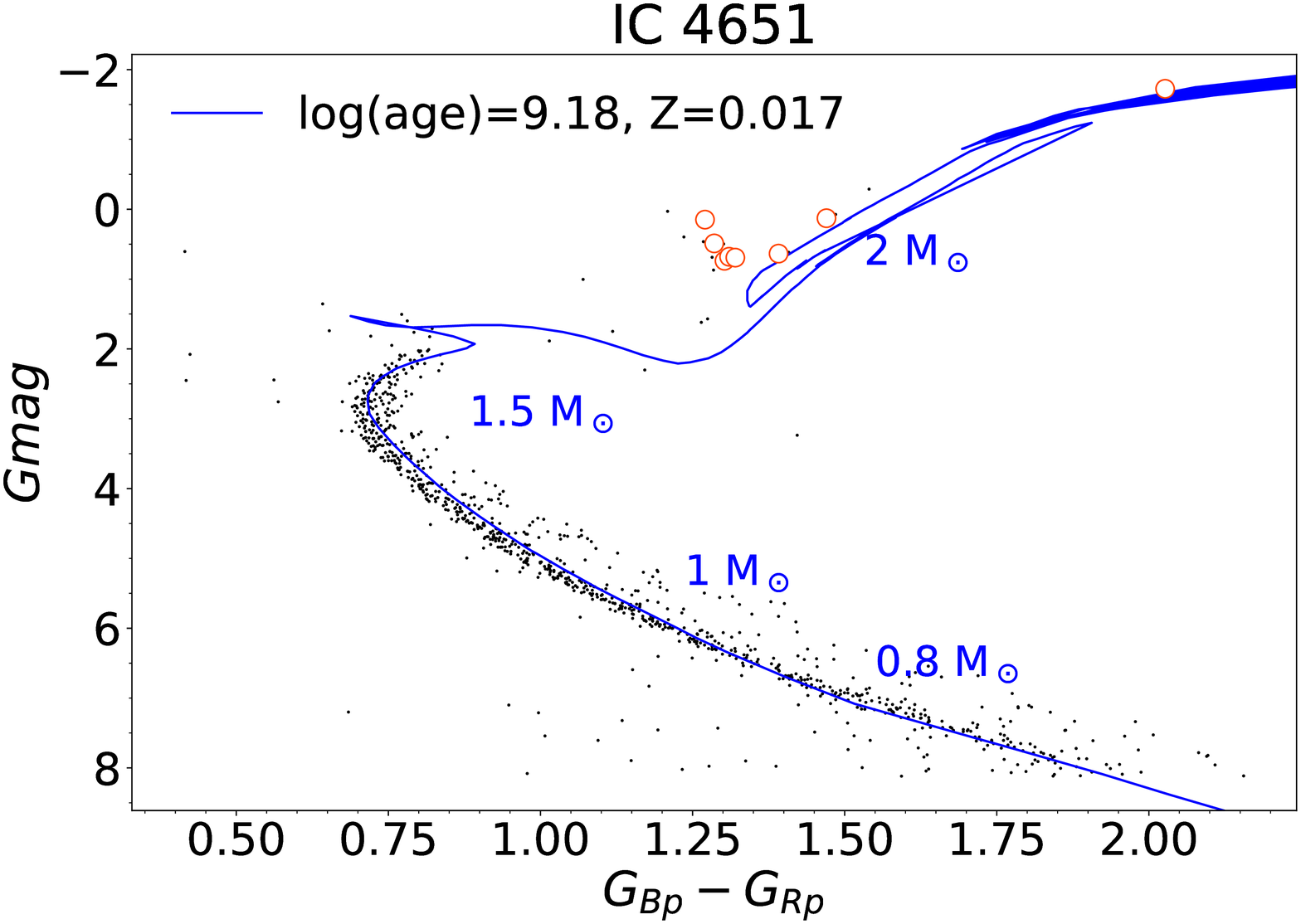}
\includegraphics[width=0.5\textwidth]{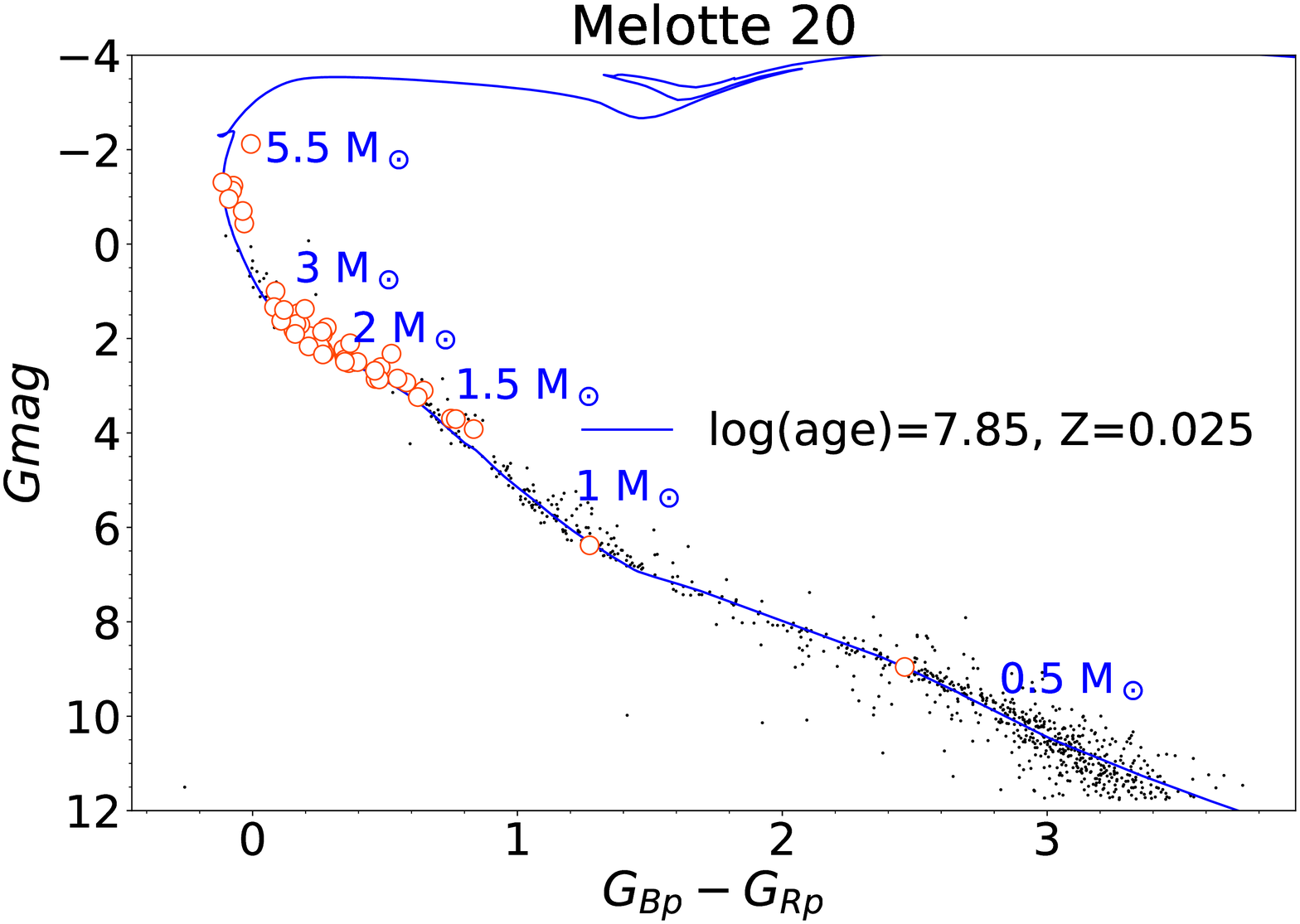}}
\centerline{\includegraphics[width=0.5\textwidth]{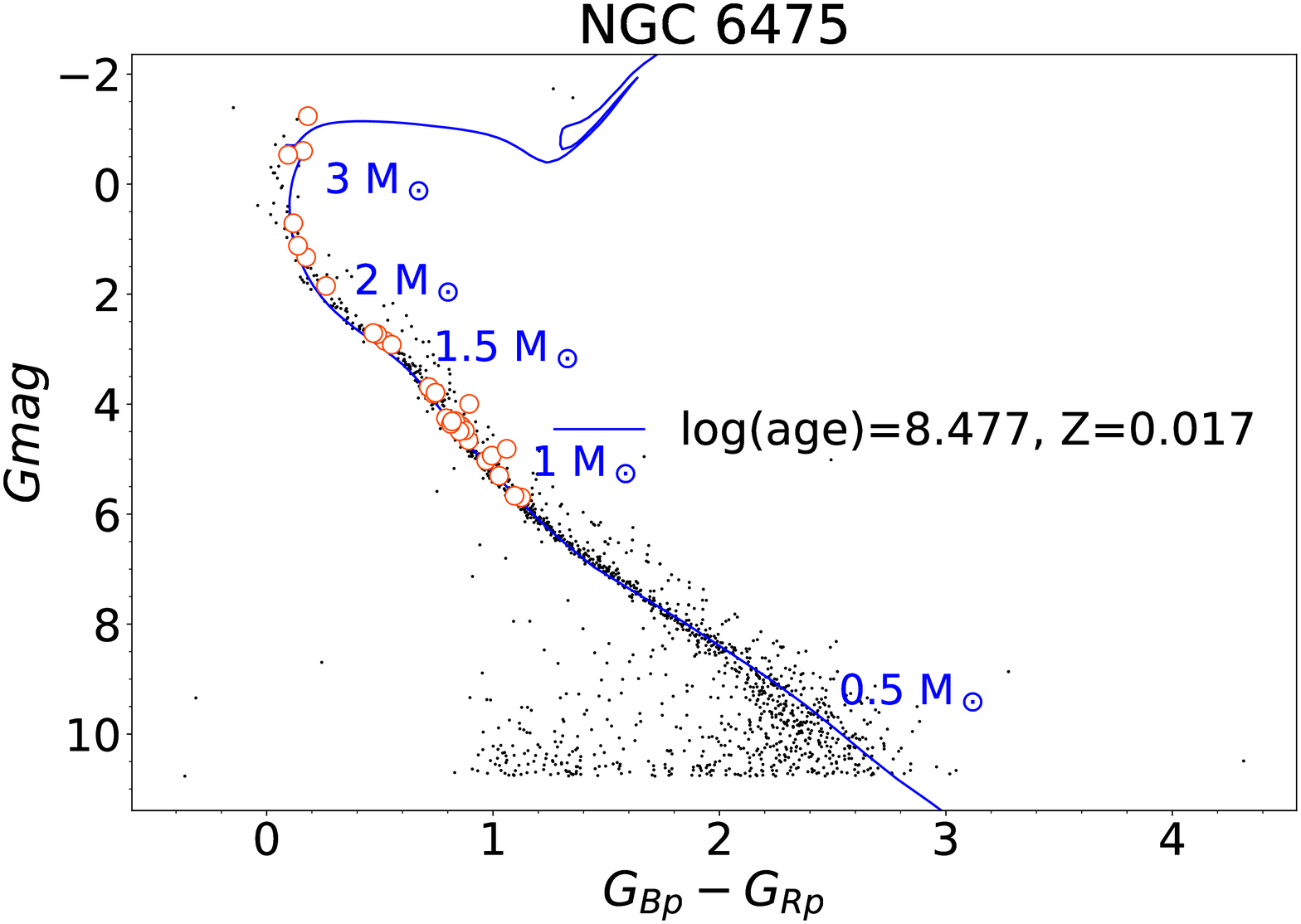}
\includegraphics[width=0.5\textwidth]{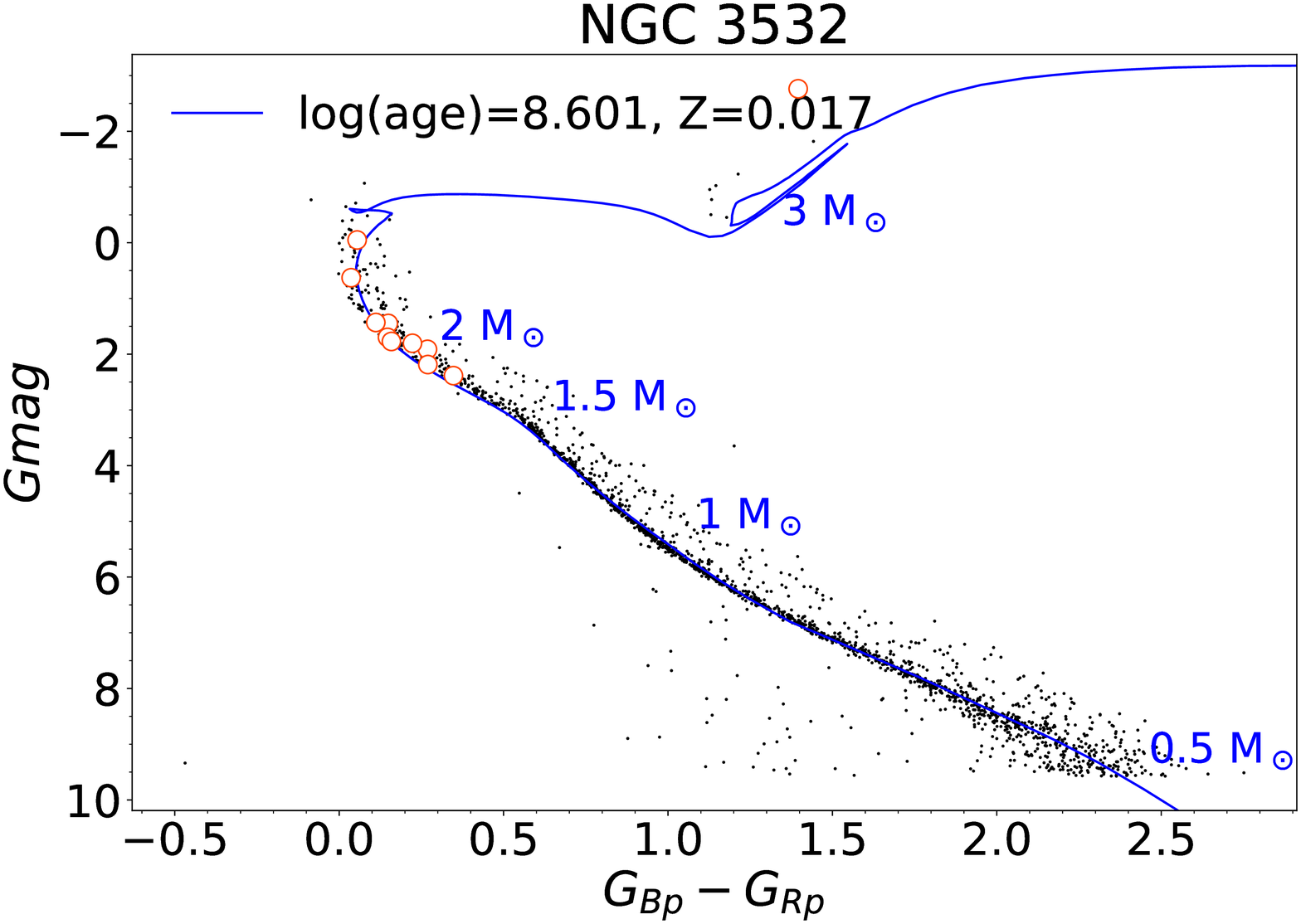}}
\caption{H--R diagram for IC\,4651, Melotte\,20, NGC\,6475 and NGC\,3532 clusters based on {\it Gaia} photometric data and compared with \texttt{PARSEC} isochrones. Red circles are TESS objects without flares.}
\label{hr_noflash}
\end{figure}

Figure~\ref{hr_flash} illustrates the Hertzsprung--Russell (H--R) diagram of several clusters from our analysis with locations of found flare stars and cluster members with {\it TESS} high temporal resolution light curves available. We also plotted the isochrones calculated using \texttt{PARSEC} \citep{PARSEC2014} for metallicities Z=0.017 and Z=0.020 updated to the latest transmission curve calibrated on {\it Gaia} DR2 data \citep{Evans2018}\footnote{\texttt{PARSEC} isochrones in {\it Gaia} DR2 passbands are available at \url{http://stev.oapd.inaf.it/cgi-bin/cmd}}.  The figure shows that the majority of the found flare stars have masses between $0.5-2~\Msun$. Figure~\ref{hr_noflash} presents several clusters where no flare star was detected. 


It is important to note that two of the clusters -- ASCC\,19 and Gulliver\,6 -- are located very close on the sky and also on similar distance. Due to this, the star HD\,290527 is considered as a member of both clusters simultaneously  \citep{2018cantat}. On the other hand, the star HD\,290674 is considered to be a member of Gulliver\,6 according  to  \citet{2018cantat} and a member of ASCC\,19 -- according to \citet{Cantat2018Tycho}. This leads to the ambiguities of the ages for these stars. 

A full list of flare stars we found is given in Table~\ref{flares}. Specified spectral types are taken from the SIMBAD database. 
As well as spectral types are not known for all star, 5th column gives effective temperatures $T_{\rm eff}$  according to {\it Gaia} DR2 \citep{Gaia2016,Gaia2018DR2}.  
The 6th column contains the number of flares detected during the period of observations, while the last column is the period of rotation. Both values were estimated using \texttt{FLATW'RM}. 
The code estimates the periods based on the strongest peak in the Lomb-Scargle periodogram, and thus may provide bogus or halved periods in some cases. Therefore we also checked all measured period values by eye and corrected them where necessary so that they clearly correspond to the main periodic components. For one of the stars in our sample -- EO\,Cha -- the determined period is close to total duration of observations and might be caused by systematic effects in the data. However, our value is in agreement with the one reported in AAVSO VSX database\footnote{American Association of Variable Star Observers Variable Star Index is available in \url{https://www.aavso.org/vsx/}}, and thus we decided to keep it in our analysis.

\begin{center}\small
\begin{longtable}[t]{ccc ccc c}
\caption{List of flare stars selected in this work. Spectral types are taken from the SIMBAD database. $T_{\rm eff}$ is temperature taken from {\it Gaia} DR2 \citep{Gaia2016,Gaia2018DR2}.  
Numbers of flares ($N_{\rm fl}$) and rotation periods ($P_{\rm rot}$) are estimated as described in Section~\ref{sec_2}.
}\label{flares}\\
\toprule
Cluster        & {\it Gaia} DR2 ID    &   Name     & Sp.      & $T_{\rm eff}$  &  $N_{\rm fl}$     &    $P_{\rm rot}$      \\
               &                &            & type     &            &               &    [days]         \\
\midrule 
\endfirsthead %
\caption{Continued.}\\
\toprule
Cluster        & {\it Gaia} DR2 ID    &   Name     & Sp.       & $T_{\rm eff}$ &  $N_{\rm fl}$     &    $P_{\rm rot}$      \\
               &                &            & type      &           &               &    [days]         \\ 
\midrule 
\endhead %
\bottomrule
\endfoot %
\bottomrule
\multicolumn{7}{l}{$^a$ HD\,290527 is probably a member of Gulliver\,6 cluster, see text for details.}
\endlastfoot
  ASCC\,19$^a$  &  3220225706694228096 &  HD\,290527                 & A3     & 7470        &  3    &   6.69  \\ 
  ASCC\,21      &  3236083864117227264 &  HD\,36030                  & B9V    & 9777        &  1    &   --  \\ 
  Alessi\,13    &  4861032719915154176 &   --                        & --     & 3922        &  1    &   2.27  \\ 
  Alessi\,13    &  4860643905115917312 &  CD-36\,1309                & F8     & 5922        &  4    &   1.20  \\ 
  Alessi\,13    &  4854771001195018752 &  CD-37\,1263                & G5     & 5637        &  5    &   2.06  \\ 
  Blanco\,1     &  2320842551136015872 &                             & K4-5V  & 4954        &  1    &   5.55  \\
  Blanco\,1     &  2320838943363481984 &  --                         & K5.3   & 4296        &  5    &   2.17  \\   
  Blanco\,1     &  2320874054720005248 &  --                         & F-K    & 4131        &  1    &   5.84  \\   
  Blanco\,1     &  2320933157765826560 & CD-30\,19826                & F-K    & 5496        &  3    &   1.03  \\   
  Blanco\,1     &  2320596123092361088 &--                           & F-K    & 5031        &  2    &   0.34   \\   
  Blanco\,1     &  2320862200611102848 & --                          & F-K    & 5499        &  7    &   2.34   \\   
  Blanco\,1     &  2320860074602561408 & --                          & F-K    & 4987        &  2    &   0.41  \\ 
  Blanco\,1     &  2320881377640416640 &--                           & F-K    & 4301        &  3    &   1.18   \\ 
  Blanco\,1     &  2320869897192498432 & CD-30\,19800                & F8     & 6070        &  1    &   2.65   \\ 
  Blanco\,1     &  2320795031617857152 &                             & F-K    & 4977        &  3    &   3.15  \\ 
  Blanco\,1     &  2320872650266808192 &  --                         & F-K    & 4873        &  3    &   4.14   \\ 
  Blanco\,1     &  2320847533296914560 & --                          & F-K    & 4579        &  4    &   6.47  \\ 
  Gulliver\,6   &  3217389348947353728 & HD\,290674                  & A0     & 9034    &  1    &   6.41   \\    
  IC\,2602      &  5239841340702856960 & HD \,93405                  & F3/5V  & 6383    &  5    &   0.75  \\  
  IC\,2602      &  5239660372284736896 &  --                         &  -     & 5362    &  4    &   3.80   \\  
  IC\,2602      &  5239626420542800512 & V570 Car                    & G9Ve   & 5071    &  2    &  1.21   \\  
  IC\,2602      &  5239498744077038976 & HD\,310131                  & G6V    & 5583    &  2    &   0.62   \\    
  IC\,2602      &  5253546997989686912 & HD\,307772                  & G7Ve   & 5161    &  3    &   0.32  \\    
  IC\,2602      &  5251470948229949568 &  --                         &  --    & 5768    &  1    &   3.02  \\    
  Mamajek\,1    &  5209082129256747008 & EG Cha                      & K4Ve   & 4365    &  8    &   4.35  \\   
  Mamajek\,1    &  5209135352491538432 & EO Cha                      & M0     & 4089    &  3    &  22.26  \\    
  Mamajek\,1    &  5209038423669475712 & EL Cha                      & M2     & 3526    &  6    &   1.84   \\    
  Mamajek\,1    &  5215178848217868288 & --                          & M3     & 3673    &  2    &   2.96   \\    
  Mamajek\,1    &  5209118305765620096 &  EQ Cha                     & M3     & 3367    & 14    &   1.25   \\    
  NGC\,1662     &  3294740954034417792 &  --                         &  --    & 4454    &  1    &   --  \\      
  NGC\,1662     &  3295557822453425792 & --                          &  --    & 4447    &  1    &  --  \\      
  NGC\,2422     &  3030122727536859008 & --                          &  --    & 4826    &  1    &   6.97  \\  
  NGC\,2451A    &  5538425822158520576 & --                          &  --    & 3761    &  2    &   0.52   \\       
  NGC\,2451B    &  5538722449779987072 & --                          &  --    & 4438    &  1    &  0.50  \\        
  NGC\,2451B    &  5538818176010943488 & --                          &  --    & 5326    &  5    &   1.95  \\        
  NGC\,2516     &  5290732721731067136 & --                          &  --    & 4882    &  2    &  2.07  \\    
  NGC\,2516     &  5290719115275116288 & --                          & A7III  & 7038    &  1    &  1.02   \\    
  NGC\,2548     &  3064487035741401984 & --                          & A2/4   & 8160    &  1    &  0.80   \\    
  NGC\,6281     &  5976330585929049600 & V948 Sco                    & ApSi   & 8553    &  1    &   1.64  \\        
  Platais\,8    &  5305106426785000448 & CPD-55\,1885                & G5V    & 5511    &  3    &  0.92  \\   
  Platais\,8    &  5303854293507991680 & HD\,78027                   & A1V    & 8521    &  1    &   3.31  \\   
  Platais\,8    &  5310127655876188544 & CD-55\,2543                 & G8V    & 5406    &  1    &   3.84  \\  
  Platais\,9    &  5427272240335502848 & --                          &  --    & 3316    &  1    &  --   \\      
  Platais\,9    &  5427469843189439360 &--                           &  --    & 3759    &  1    &   3.49   \\  
  Platais\,9    &  5423893028786327168 & HD\,80484                   & A1V    & 8884    &  1    &  0.69  \\  
  Platais\,9    &  5327362504934951040 &                             &  --    & 5732    &  1    &   1.23   \\   
  Melotte\,111  &  3959841756787187456 &        --                   & M3.0   & 3395    &  2    &   --  \\
  Melotte\,111  &  4002505586787874560 &      --                     & M3.5   & 3421    &  5    &  --  \\
  Melotte\,111  &  3960130997064687616 &    --                       & M3.5   & 3783    &  5    &  --  \\
  Melotte\,111  &  4008440617411511552 &    Sand\,64                 & M3.9   & 3921    &  2    &  --  \\
  Melotte\,111  &  4009049575054518400 &    --                       & M2.4   & 3789    &  1    &  --  \\
  Melotte\,111  &  3954074577781239936 &    Sand\,63                 & M2.2   & 3628    &  2    &   --  \\
  Melotte\,111  &  4009398051520907008 &    HK Com                   & M4.2   & 3848    &  4    &  --  \\
  Melotte\,111  &  4009440382718550272 &    --                       & M3.7   & 3937    &  1    &  --  \\
  Melotte\,111  &  4009455054326835968 &      --                     & M2.6   & 3899    &  3    &  3.39  \\
  Melotte\,111  &  4002543008838439296 &      --                     & M3.8   & --      &  1    &  1.87   \\
\end{longtable}
\end{center}

\section{Discussion}\label{sec_3}
\subsection{Spectral distribution of selected flare stars}

\begin{figure}
\centerline{
\includegraphics[width=0.5\textwidth]{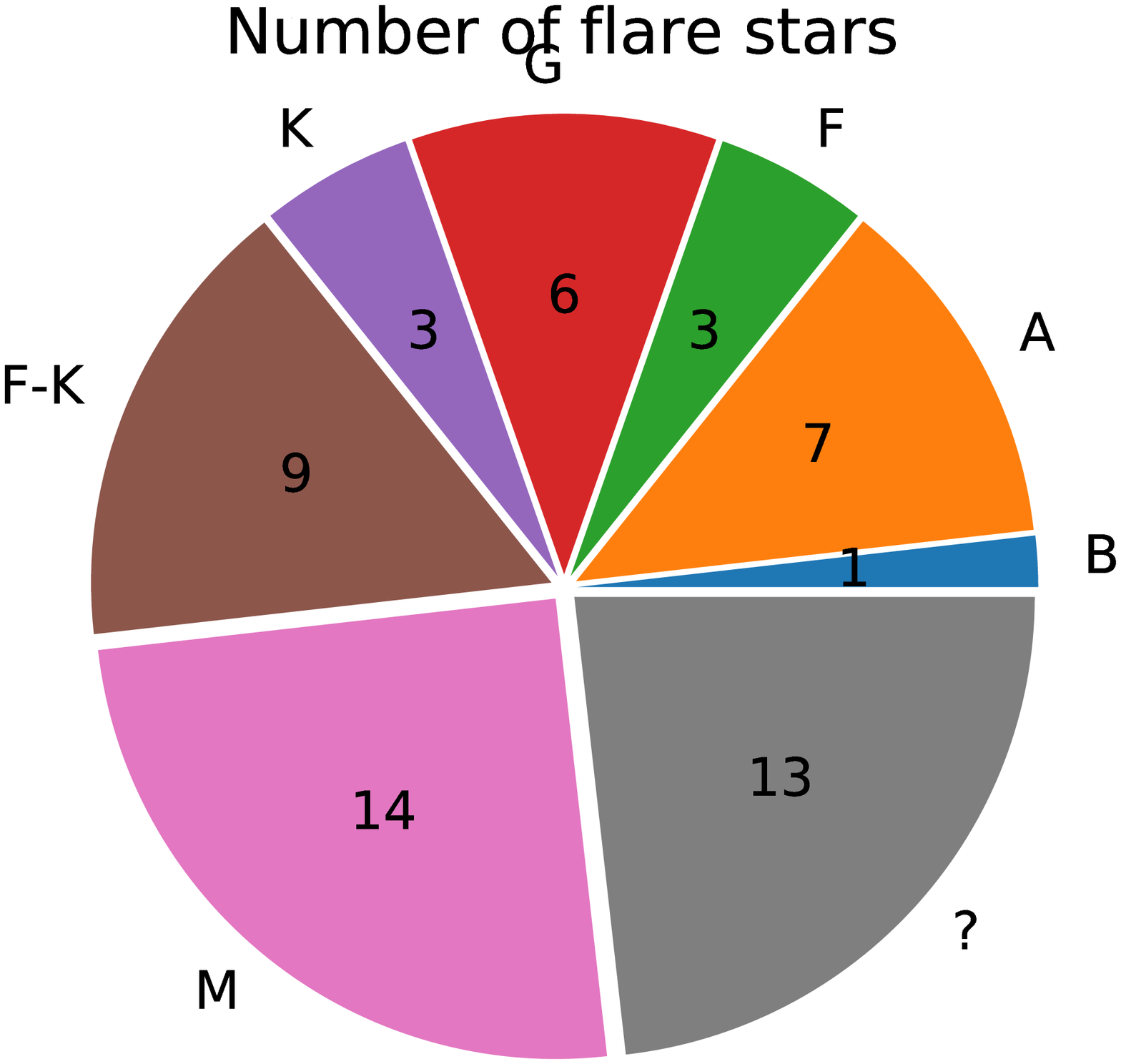}
\includegraphics[width=0.5\textwidth]{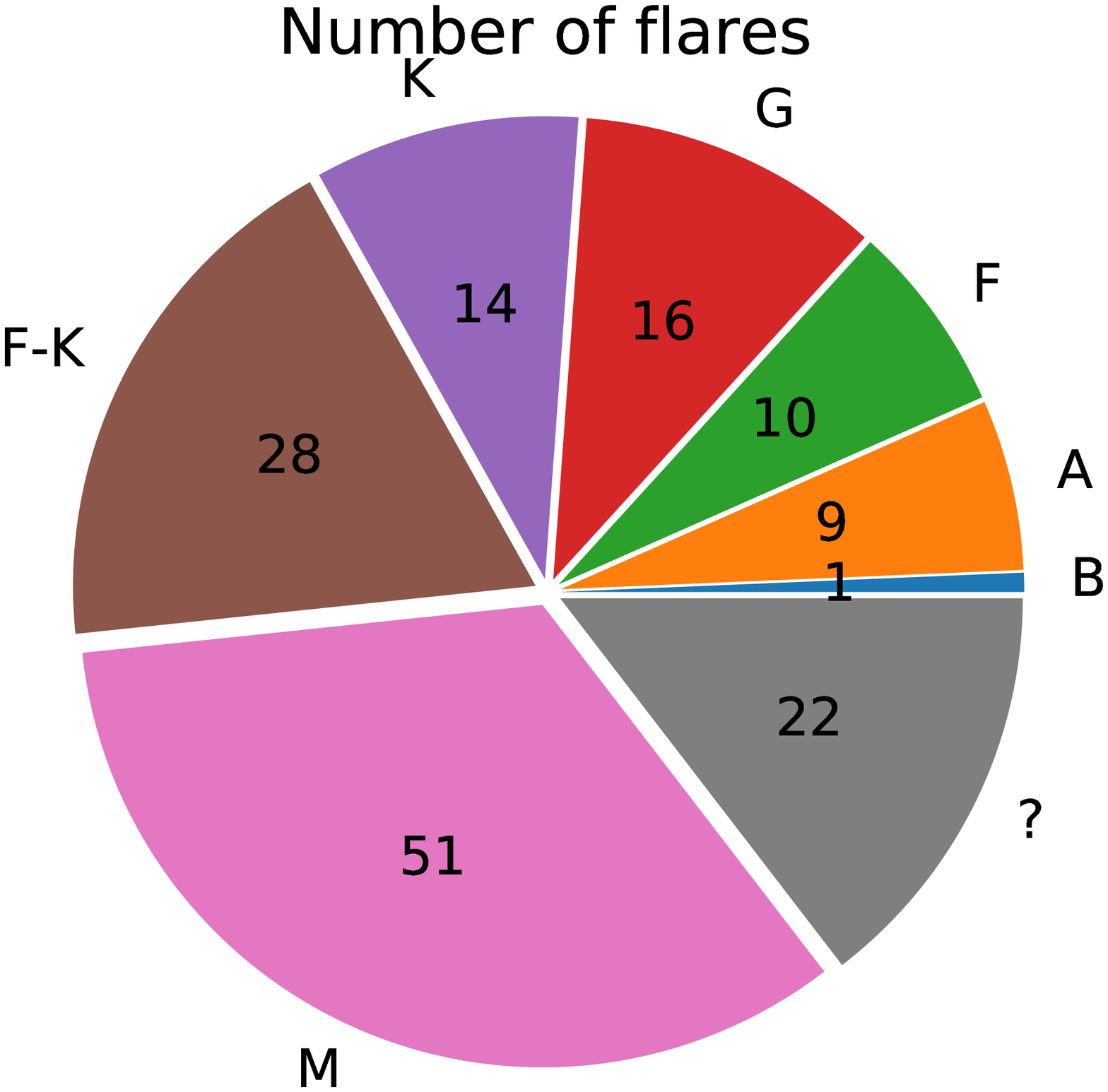}
}
\caption{Distribution of flare stars in our sample (left panel), and detected flares from them (right panel) over spectral types from Table~\ref{flares}.} 
\label{sptypeflare}
\end{figure}

Most of stellar flares occur in red dwarfs, more often in cool M dwarfs \citep{gudel2009}. Data from the first data release of {\it TESS} mission confirm it: among 1228 flare stars detected by \citet{gunther2020}, 673 objects were classified as M dwarfs. Our data is in a good agreement with these previous studies. Figure~\ref{sptypeflare} presents the histogram of the number of flare stars versus spectral type based on the data from Table~\ref{flares}. Indeed, the majority of selected objects are stars of late spectral type. 

\begin{figure}
\centerline{\includegraphics[width=0.5\textwidth]{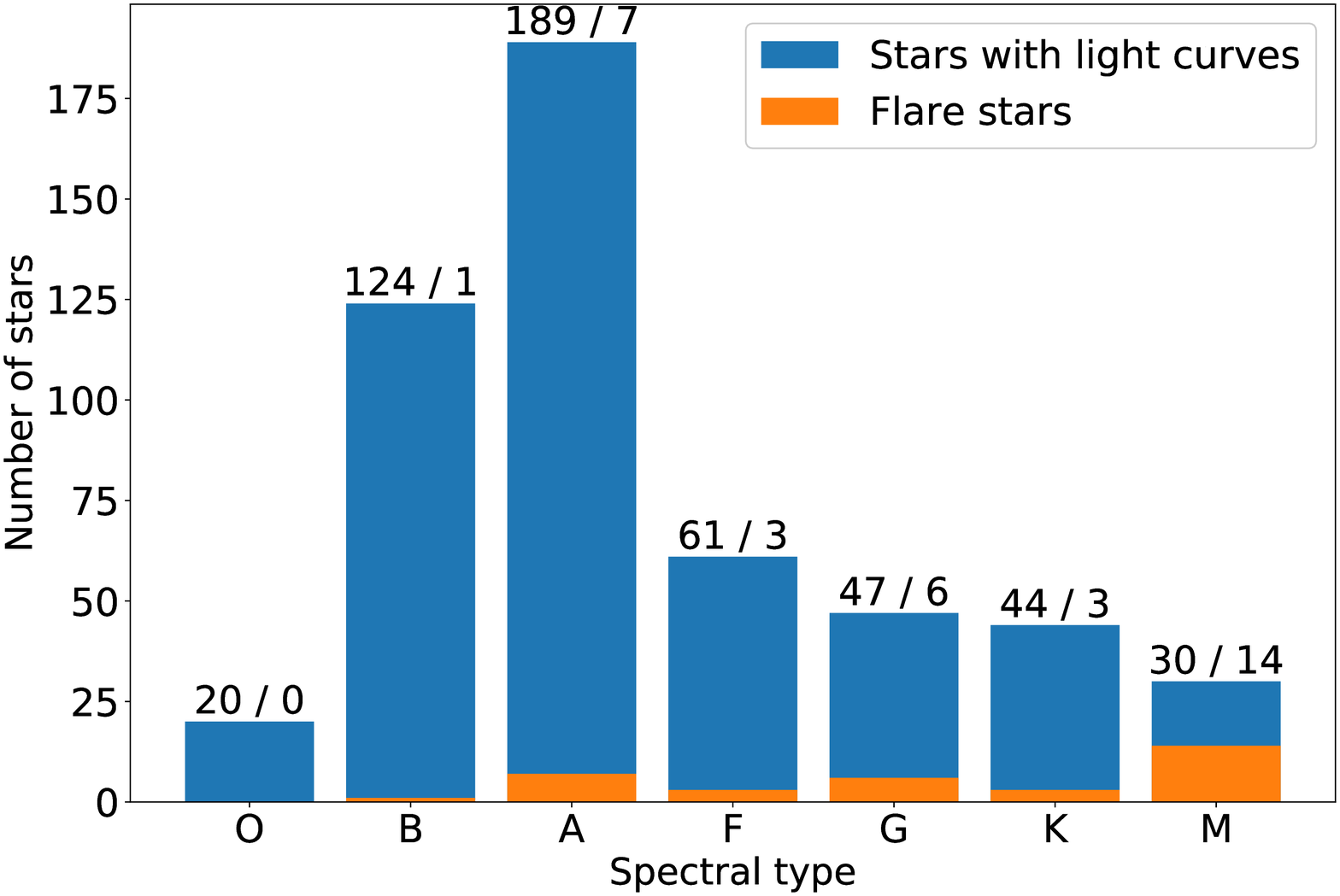}
\includegraphics[width=0.5\textwidth]{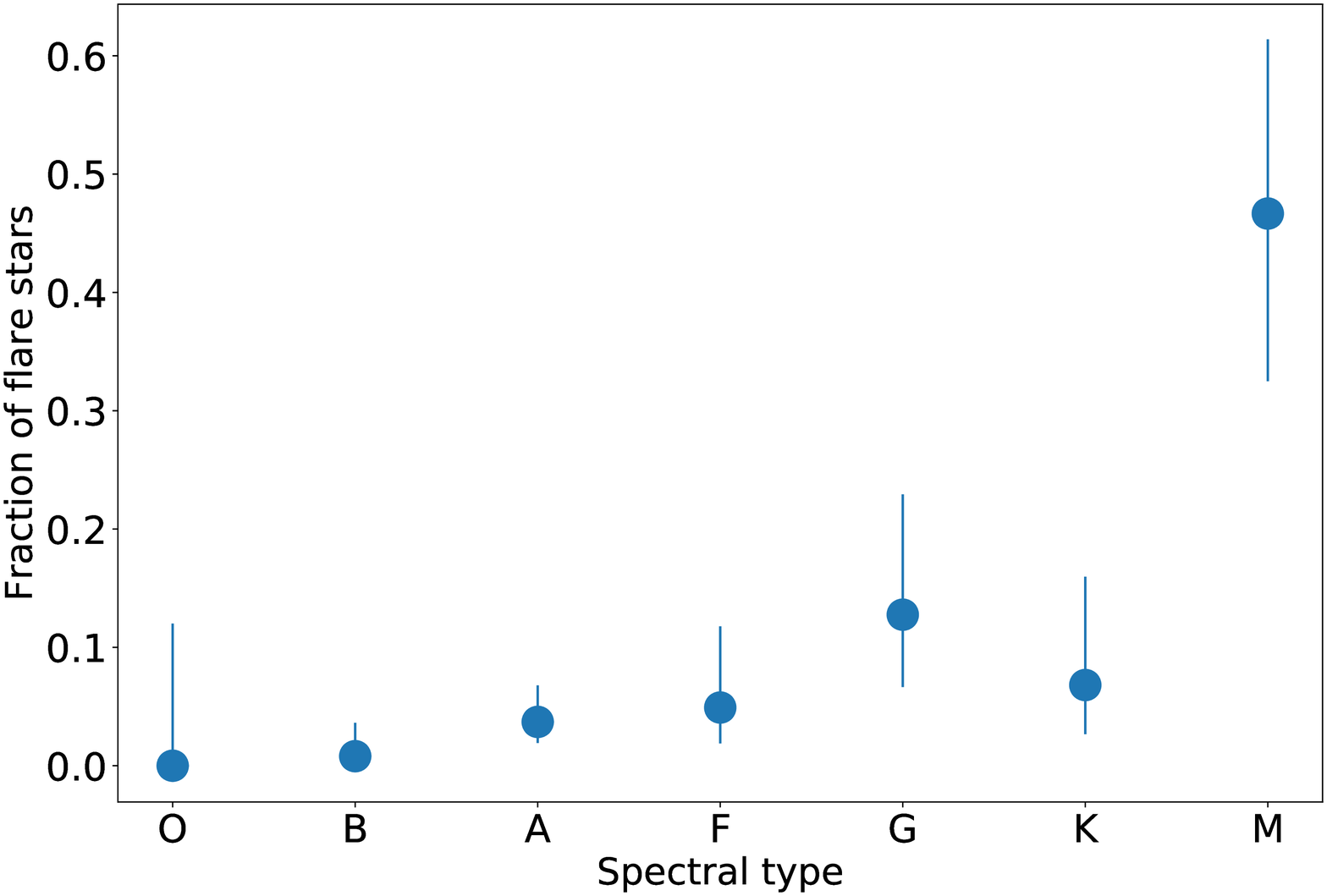}}
\caption{Left panel -- histogram of total number of light curves and flare stars among them per spectral type (where it was reliably determined). \newline
Right panel -- fraction of flare stars as a function of spectral type. Error bars depict the 90\,\% confidence intervals for the corresponding values.}
\label{sptypefrac}
\end{figure}

M-type dwarfs are very faint  objects, an absolute visual magnitude of M0V star is $M_{V} = 9.0$~mag and for M5V is $M_{V} = 14.6$~mag \citep{Schmidtkaler}. In our work, we found M-type flare stars only in two nearby open clusters, Mamajek\,1 (distance $D$~=~98.72\,pc) and Mel\,111 ($D$~=~96\,pc). Figure~\ref{sptypefrac} shows that flares were detected for almost 50\,\% of all considered M-type stars.

Stellar flares have been previously detected in some hot B--A type stars (in optical range \citep{Schaefer1989,balona2012MNRAS}; in X-ray \citep{schmitt1994, yanagida2007}). 
Several objects from our sample are also early spectral type flare stars. In this study, we detected flares in seven A-type stars (Table~\ref{flares}). One of them -- V948\,Sco  -- belongs to $\alpha^2$ CVn variables, chemically peculiar main sequence stars with a strong magnetic field. For all other A-type stars, the variability was registered for the first time.  
\begin{figure}
\centerline{\includegraphics[width=0.77\textwidth]{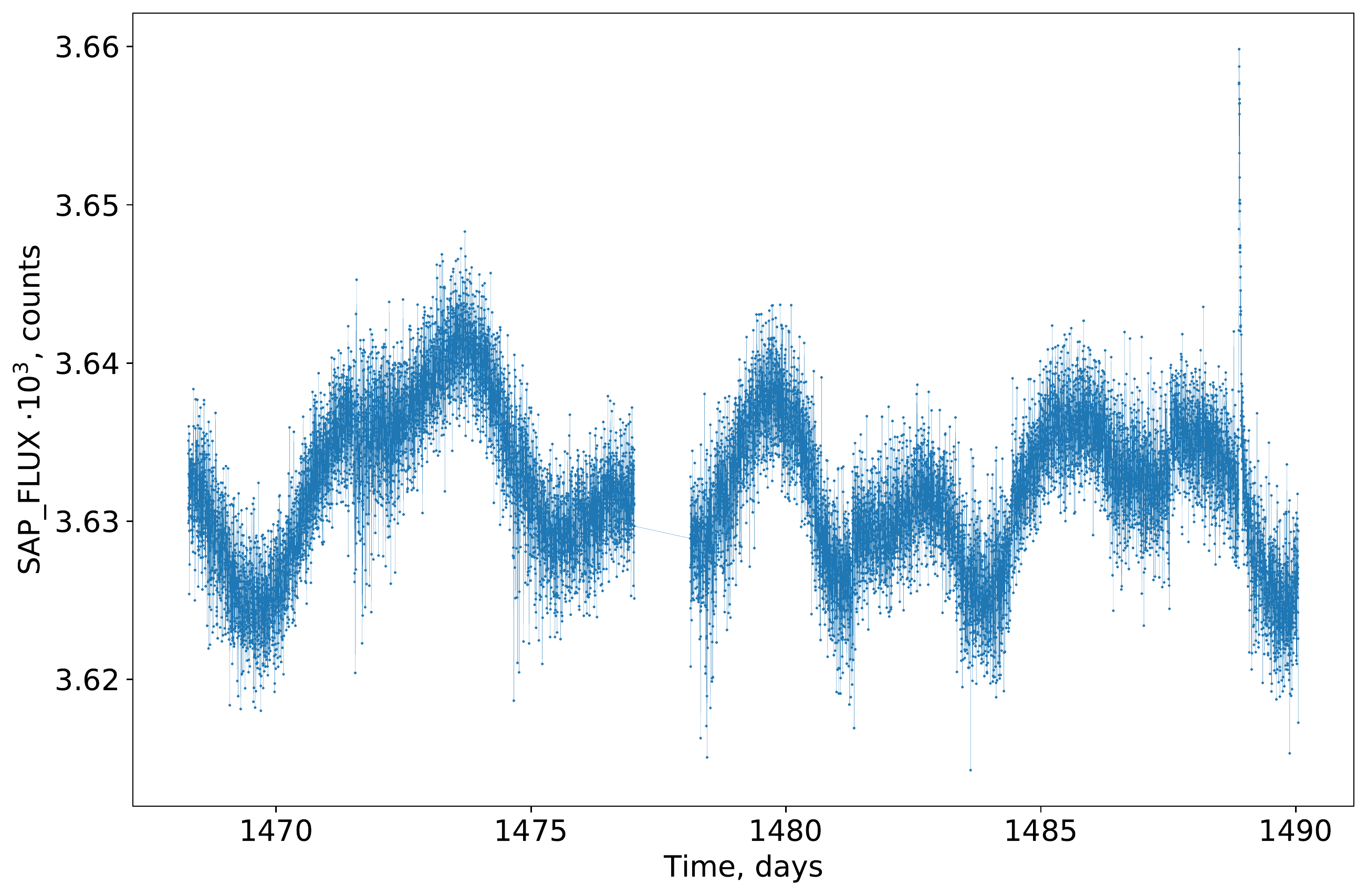}
\includegraphics[width=0.23\textwidth]{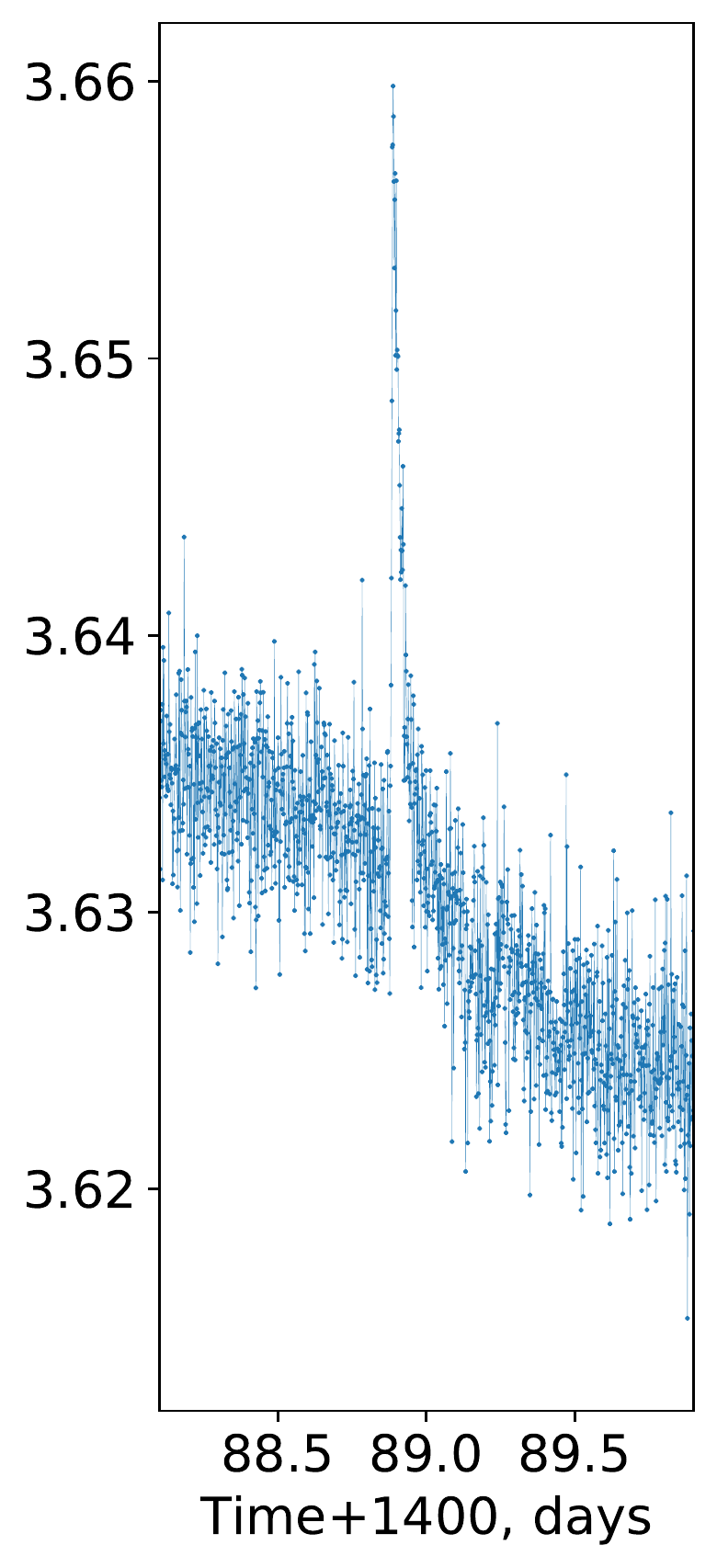}}
\caption{Light curve of HD\,36030 (B9V) with a clearly detected flare.}
\label{hd36030}
\end{figure}

Another selected hot star is HD\,36030, classified as B9V star by \citet{HoukSwift1999}. Figure~\ref{hd36030} presents the light curve of HD\,36030 and a clearly detected flare. As in the case with A-type stars, we did not find any previously published work about variability of this object. 

Flares of hot stars may be explained through binarity. As mostly A-type stars are probably double or multiple systems, it is natural to expect flares to originate in a cool companion. However, \citet{balona2019} gives strong arguments that flares originate in A-type stars themselves. These arguments are based on ratio of luminosity: an M- or K-type companion is about 50--100 times less luminous than an A-type star. The flare amplitude originating in a cool companion should therefore, on average, appears to be about 50--100 times smaller when observed together with A-type star. However, flares detected in hot stars are comparable with their luminosity \citep{balona2019}. 
Flares in A-stars of our sample have amplitudes of couple percents only, and may be in principle, at least partially, explained by a giant flares with amplitudes of several magnitudes on a late type companion, like the ones reported in \citet{Beskin2017}. However, such giant flares are quite rare, and are not detected in our sample of M stars. Thus, we may not expect them to be abundant among the possible components of our A stars too.
On the other hand, magnetic interaction with the companion star could provide a plausible explanation for observed flaring activity \citep{balona2012MNRAS}.

According to \citet{balona2015}, the relative number of flare stars from cool M dwarfs to hot A-type stars is probably the same. According to our results (see Figure~\ref{sptypefrac}), 
the fraction of flare stars among G-M type objects is significantly higher.


 
There are already no published data about spectral types for 
fourteen flare stars in our sample (see Table~\ref{flares}, Figure~\ref{sptypeflare}). Future spectral observations of these objects could improve statistics. 

\subsection{Statistical analysis of flares }

The mean duration of flares for stars in Table~\ref{flares} varies between 20 to 70 minutes. This timescale is comparable with the one of GOES X-class solar flares, quite often visible in the white light \citep{Harra2016}. Only 4 stars among 56 analysed in this paper have more than 5 flares during period of {\it TESS} observations. One of them is EQ\,Cha (M3) -- variable star of Orion type, close visual binary \citep{Sicilia2009} and a member of Mamajek\,1 open cluster. It showed 14 flares during 25 days of observations. Flaring activity of EQ\,Cha star has been already reported in X-ray by \citet{EQCha2010}. 

\begin{figure}
\centerline{
\includegraphics[width=0.5\textwidth]{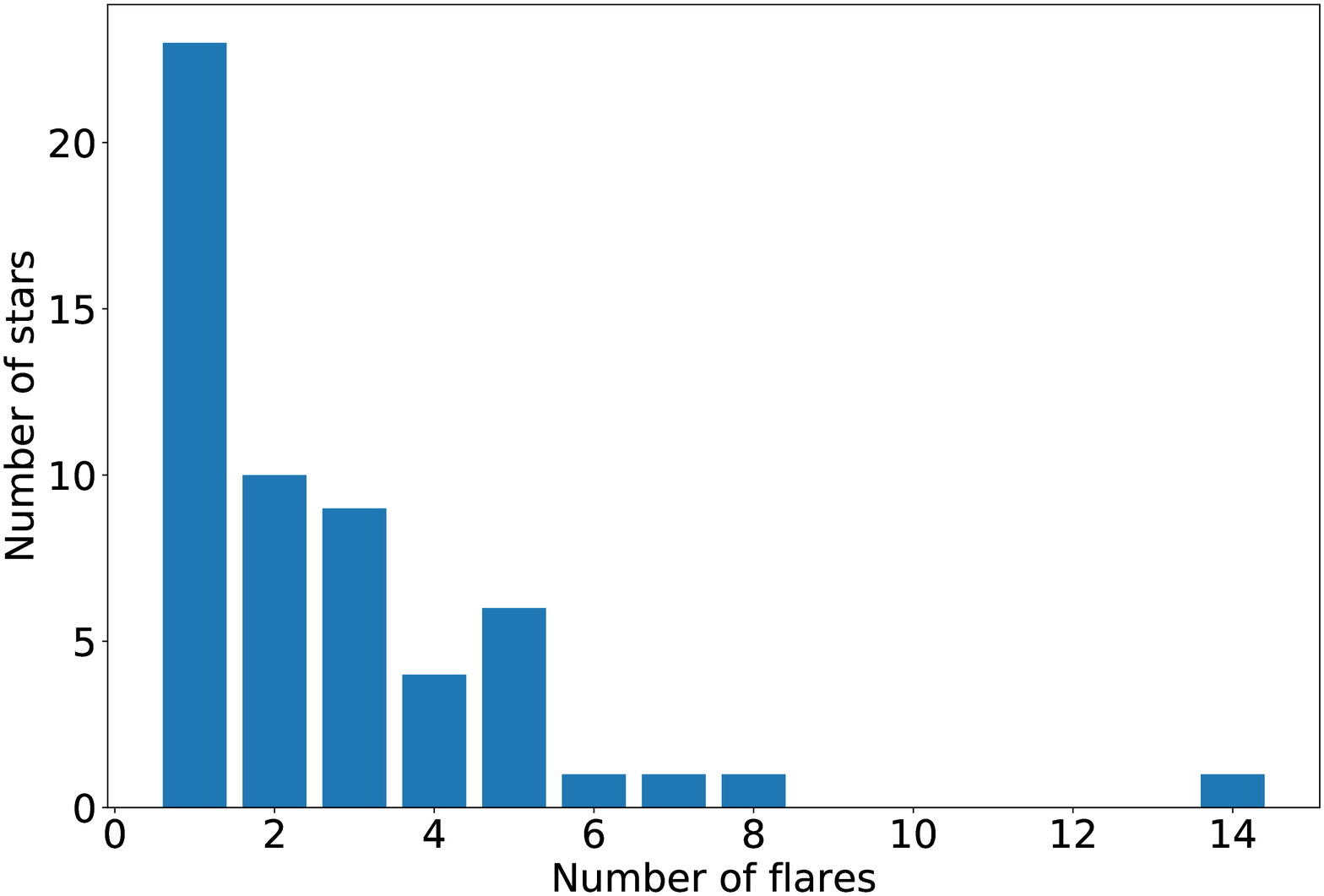}
\includegraphics[width=0.5\textwidth]{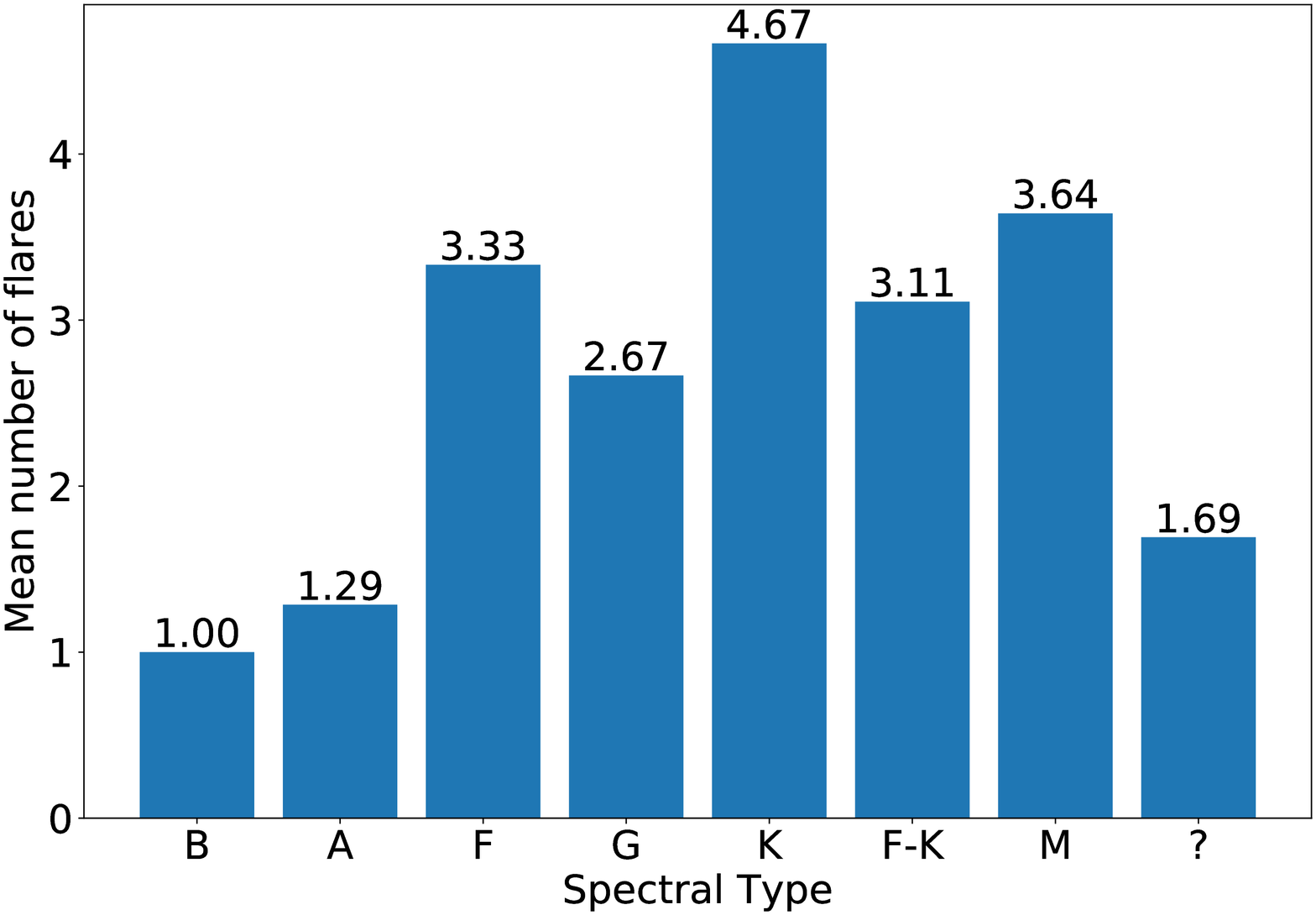}
}
\caption{Histogram of number of flares detected from a star (left panel). Mean number of flares per star depending on spectral class from Table~\ref{flares} (right panel).}
\label{total}
\end{figure}
\begin{figure}
\centerline{\includegraphics[width=1.0\textwidth]{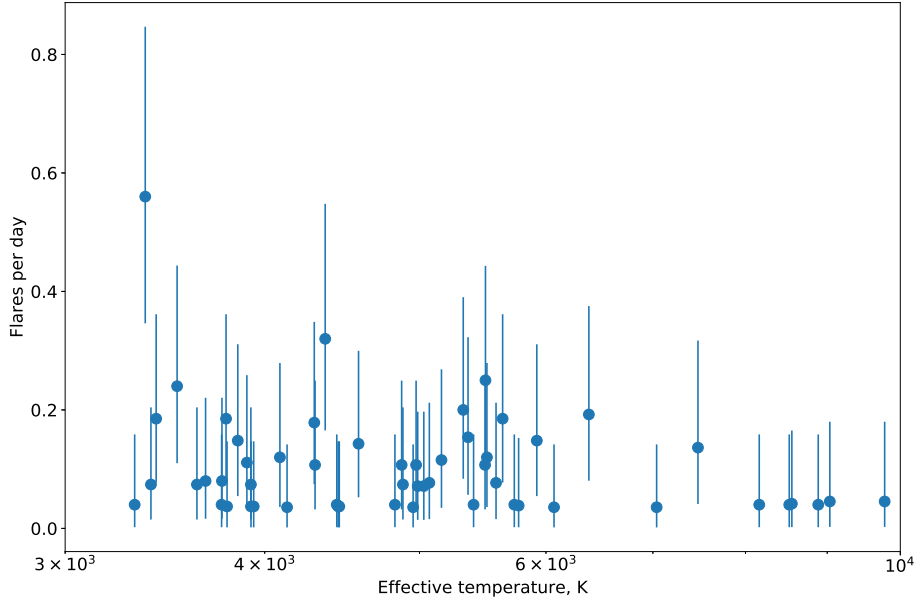}}
\caption{Rate of stellar flares as a function of temperature. Error bars correspond to the 90\,\% confidence intervals computed assuming Poissonian statistics.}
\label{teff}
\end{figure}
\begin{figure}
\centerline{\includegraphics[width=1.0\textwidth]{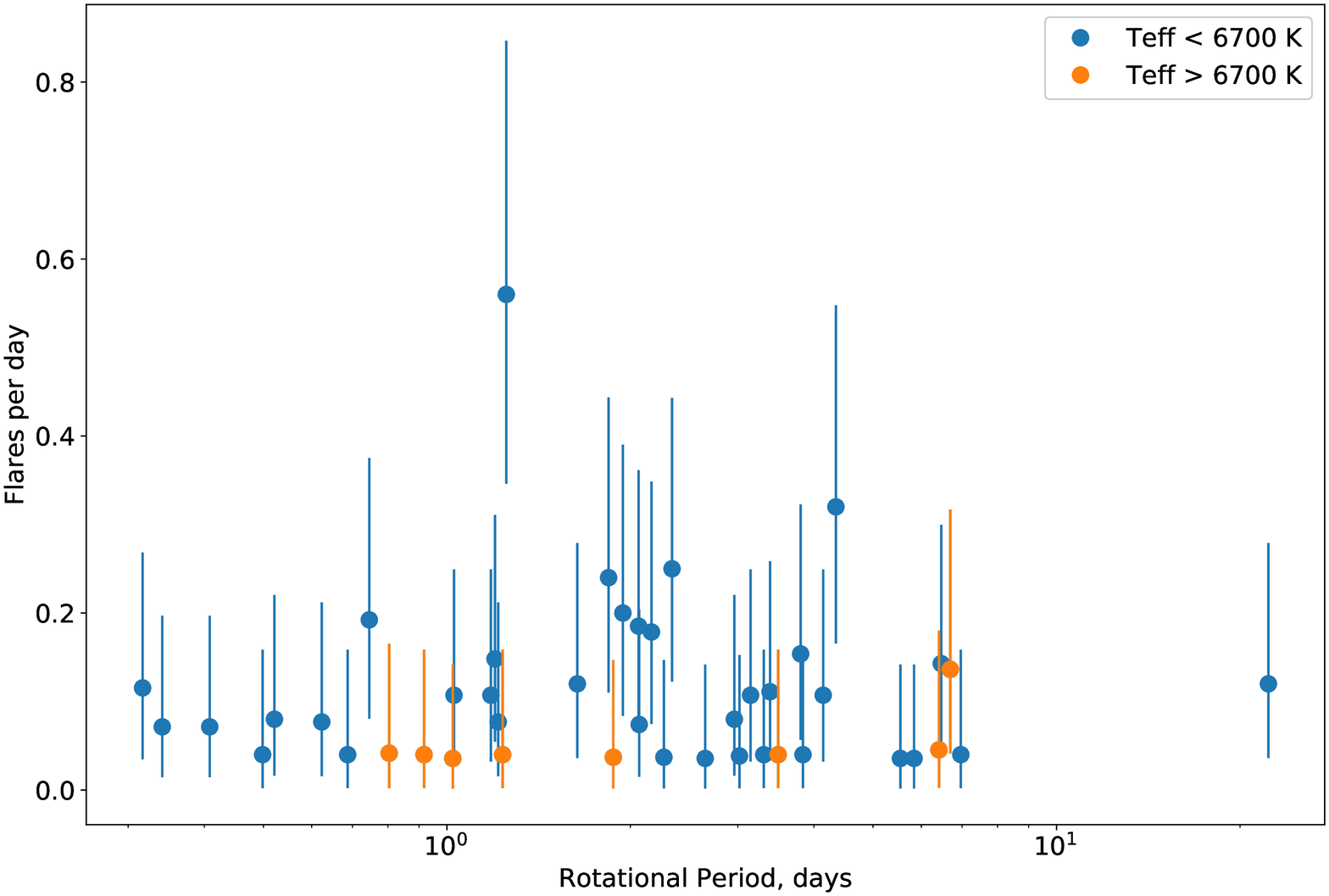}}
\caption{Rate of stellar flares as a function of rotational period. Error bars correspond to the 90\,\% confidence intervals computed assuming Poissonian statistics.}
\label{rotation}
\end{figure}

On the right panel of Figure~\ref{total}, we can see that K-type stars have the largest mean flare activity in their light curves. 
The largest number of flares belongs to M-type stars, which are all earlier than M4 in our sample (see Table~\ref{flares}) and thus are cool stars with convective envelope, as well as to K-type stars. 
These K and early M dwarfs, which retain a radiative core, presumably all have a~quite efficient solar-type dynamo  powering  their magnetic activity. On the other hand, M dwarfs of spectral type later than M4 are fully convective, and have a turbulent dynamo at work \citep{Wright2011, Stassun2011}.
The activity of G-type stars could be also associated with their convective envelopes. 
Figure~\ref{teff} shows the dependence of a flare rate on the star' effective temperature. 

The appearance of spots on the surface of stars and stellar flares are clear evidences of stellar magnetism. The connection between stellar rotation and magnetic activity has been studied in many works. For example, \citet{Guo2014} analyzed X-class flares of Sun during the 22nd and 23rd solar cycles, and they found that flares closely follow the same 11 year cycle as sunspots. \citet{doyle2018} did not find any correlation between the rotation  phase and the number of flares for M-dwarfs based on {\it Kepler} data, but clearly detected the decrease of flaring activity for rotational periods longer than 10 days. The same dependence is seen in X-ray activity-rotation relation \citep{Pizzocaro2019}.

Figure~\ref{rotation} shows the dependence of flaring activity in our sample, estimated as a number of flares per day of observations on the rotational period of the star. We may spot just a slight trend of decreasing activity towards slower rotation. While we may not stress it as a significant dependence, let's note that it agrees with the expected behaviour of magnetic activity being as a result of internal magnetic dynamo, arising from the combination of stellar differential rotation and convection in the sub-photospheric layers.

\subsection{Age distribution of flare stars} 

\begin{figure}
\centerline{\includegraphics[width=1.0\textwidth]{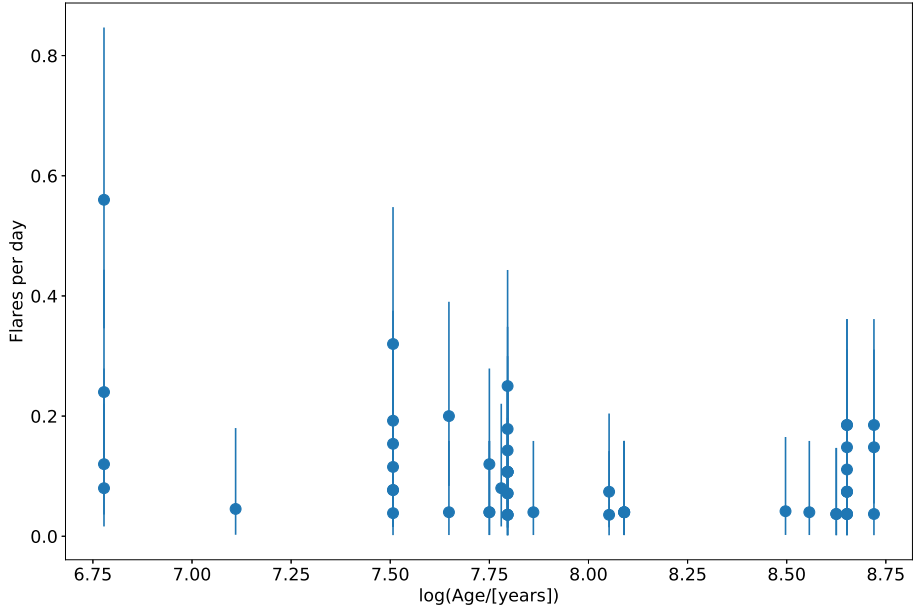}}
\caption{Rate of flares as a function of age. The rate was estimated from the number of detected flares divided by the effective light curve duration for every individual star. Error bars depict the 90\,\% confidence intervals computed assuming Poissonian statistics.}
\label{agerate}
\end{figure}

Mamajek\,1 is the youngest open cluster among ones where we have found flare stars. Its age is 6~Myr. The oldest one where we have found flare is the cluster NGC\,1662. Its age 1\,Gyr is comparable with the maximal lifetime of open clusters \citep{Fujii2016}. The ages of all examined clusters span from 1~Myr (IC\,5146, the youngest) to 4.3~Gyr (NGC\,188, the oldest one). Therefore, the collected data do not allow us to confidently determine the age after which flare stars in clusters disappear. 

Figure~\ref{agerate} shows how flaring activity changes with stellar age. 
The resulting distribution is mostly flat, with just a single deviating point with 14 detected flares from EQ\,Cha star. We therefore cannot reliably see any dependence of flaring activity on the age.

\section{Conclusion}\label{sec_4}

In order to study the dependence of stellar flaring activity on the age, we performed an uniform study of large sample of Galactic open clusters from catalogue of \citet{2018cantat}. We analyzed a total amount of 957 high-cadence {\it TESS} light curves for stars from 136 open clusters. By visual inspection and later using a \texttt{FLATW'RM} code, it finally turned out that 56 of them are flare stars with 151 flares detected.

41.5\,\% of all flares were detected in cool stars. 8 flare stars belong to B- and A-type stars. The flare activity was detected for the first time for all these hot stars. The flare with the largest amplitude appears on M3-type EQ\,Cha star. From total amount of detected flare stars in our sample, 25\,\% of them are members of spectral type M, 11\,\% are G-type stars and about 5.5\% F- and K-type stars. 
\newline
Statistical analysis of detected flare stars shows that:
\begin{itemize}
    \item among cool stars up to 50\,\% have flares, while among A-type stars it is only 3.7\,\%.
    \item duration of flares varies between 20 to 70 minutes.  
    \item we found no significant link between rotation rate and flare activity.
    \item we found no significant link between stellar age and flare activity.
\end{itemize}

In the study, we used {\it TESS} data recorded at 2-minute cadence light curves, which are an invaluable source of photometric data for various kinds of stars, pre-selected before the mission start. Although {\it TESS} full-frame images have 
a much wider coverage of objects, limited only by crowding, their utility for studies of stellar flares are significantly lower due to much worse temporal resolution (30 minutes, which is comparable to total duration of flares). As TESS mission now continues into its third year, re-visiting the regions of the sky already observed before, we expect to get more data for the objects we studied in the future. In the future, we plan to do spectroscopic observations of  B-A type stars with flare activity we had detected, in order to investigate their possible multiplicity.

\acknowledgements
The authors are extremely thankful to organizers of GATE Summer School for the idea of this project. 
The authors acknowledge the support from ERASMUS+ grant number 2017-1-CZ01-KA203-035562. We are grateful to the anonymous referee for useful comments and suggestions on the manuscript. %
O.M. acknowledges support from the Czech Science Foundation GA18-05665S. K.V. acknowledges the support of the Lend\"ulet Program  of the Hungarian Academy of Sciences, project No. LP2018-7/2019, the NKFI KH-130526 and NKFI K-131508 grants.

This  paper  includes  data  collected  by  the  TESS  mission,  which  are  publicly  available  from  the  MikulskiArchive for Space Telescopes (MAST). Funding  for  the  TESS  mission is  provided  by  NASA's  Science  Mission  directorate. 
This research was made by using of the SIMBAD data base and the VizieR catalogue access tool, both operated at CDS, Strasbourg, France. The WEBDA database, operated at the Department of Theoretical Physics and Astrophysics of the Masaryk University, and data from the European Space Agency (ESA) mission {\it Gaia}\footnote{\url{https://www.cosmos.esa.int/gaia}}, processed by the {\it Gaia} Data Processing and Analysis Consortium (DPAC\footnote{\url{https://www.cosmos.esa.int/web/gaia/dpac/consortium}}). 
Funding for the DPAC has been provided by national institutions, in particular the institutions participating in the {\it Gaia} Multilateral Agreement. 

\bibliography{flares}

\clearpage

\appendix{Clusters with {\it TESS} light curves available }
\begin{center}
\small
\begin{longtable}[t]{lcc ccc r}
\caption{Complete list of Galactic open clusters investigated in this work, with ages $\tau$ taken from various sources (see the last column),
distances $D$ and number of members $N_{\rm{mem}}$ taken from \citet{2018cantat}, number of available light curves $N_{\rm{lc}}$ in the {\it TESS} database and number of flare stars $N_{\rm{fs}}$ estimated in this work.}
\label{OCs_list}\\ %
\toprule
\multicolumn{1}{l}{Name} & \multicolumn{1}{c}{log($\tau$/[yr])} & \multicolumn{1}{c}{$D$ [pc]} & \multicolumn{1}{c}{$N_{\rm{mem}}$} & \multicolumn{1}{c}{$N_{\rm{lc}}$}  & \multicolumn{1}{c}{$N_{\rm{fs}}$} & \multicolumn{1}{r}{Reference for age}\\ 
\midrule 
\endfirsthead %
\caption{Continued.}\\ %
\toprule
\multicolumn{1}{l}{Name} & \multicolumn{1}{c}{log($\tau$/[yr])} & \multicolumn{1}{c}{D [pc]} & \multicolumn{1}{c}{N$_{\rm{mem}}$} & \multicolumn{1}{c}{N$_{\rm{lc}}$}  & \multicolumn{1}{c}{N$_{\rm{fs}}$} & \multicolumn{1}{r}{Reference for age}\\ 
\midrule  
\endhead %
\bottomrule
\endfoot %
\bottomrule
\multicolumn{7}{l}{$^a$ WEBDA database https://webda.physics.muni.cz/}\\
\multicolumn{7}{l}{$^b$ Distance for Melotte\,111 are taken from WEBDA.}
\endlastfoot
Alessi 13 	     &  8.72   & 104.34 	        & 48 	&     10  & 3  &   \cite{2018yen} \\
Alessi 19 	     &  7.38   & 584.80 	        & 74 	&     1   & 0  &   \cite{2019bossini}\\
Alessi 20 	     &  8.22   & 432.71 	        & 124 	&     2   & 0  &    WEBDA$^a$ \\
Alessi 21 	     &  7.816  & 581.40 	        & 137 	&     2   & 0  &   \cite{2019bossini} \\
Alessi 24 	     &  7.945  & 485.67 	        & 163 	&     1   & 0  &   \cite{2019bossini} \\
Alessi 3 	     &  8.90   & 280.43 	        & 178 	&     8   & 0  &   \cite{2018yen} \\
Alessi 37 	     &  8.125  & 722.54 	        & 142 	&     1   & 0  &   \cite{2019bossini} \\
Alessi 43 	     &  7.48   & 968.99 	        & 318 	&     2   & 0  &    WEBDA\\
Alessi 5 	     &  7.723  & 399.84 	        & 302 	&     3   & 0  &   \cite{2019bossini}\\
Alessi 9 	     &  8.42   & 207.38 	        & 194 	&     5   & 0  &   \cite{2018yen} \\
Alessi Teutsch 12&  7.977  & 612.75 	        & 44 	&     1   &  0 &   \cite{2019bossini} \\
Alessi Teutsch 5 &  7.02   & 900.09 	        & 158 	&     1   &  0 &    WEBDA \\
Alessi Teutsch 8 &  8.65   & 1034.13 	        & 341 	&     2   &  0 &    WEBDA \\
ASCC 105 	     &  7.994  & 560.85 	        & 127 	&     1   &  0 &   \cite{2019bossini} \\
ASCC 10 	     &  8.599  & 685.40 	        & 71 	&     1   &  0 &   \cite{2019bossini} \\
ASCC 12 	     &  8.42   & 1062.70 	        & 162 	&     1   &  0 &    WEBDA\\
ASCC 16 	     &  7.047  & 352.36 	        & 226 	&     9   &  0 &  \cite{2019bossini} \\
ASCC 19 	     &  7.086  & 361.27 	        & 188 	&     3   &  1 &  \cite{2019bossini} \\
ASCC 21 	     &  7.032  & 348.92 	        & 131 	&     3   &  1 &  \cite{2019bossini} \\
ASCC 32 	     &  7.404  & 813.01 	        & 259 	&     2   & 0  &  \cite{2019bossini} \\
ASCC 58 	     &  7.826  & 485.44 	        & 137 	&     2   & 0  &  \cite{2019bossini}\\
ASCC 79 	     &  6.86   & 850.34 	        & 129 	&     1   & 0  &  \cite{2019bossini} \\
ASCC 85 	     &  7.42   & 893.66 	        & 119 	&     1   & 0  &    WEBDA \\
Aveni Hunter 1 	 &  8.26   & 425.89 	        & 82 	&     3   & 0  &    WEBDA \\
Basel 8 	     &  8.102  & 1600.00 	        & 33 	&     1   & 0  &    WEBDA \\
Berkeley 86 	 &  7.116  & 1792.11 	        & 31 	&     2   & 0  &    WEBDA \\
Berkeley 87 	 &  7.152  & 1745.20 	        & 131 	&     1   & 0  &    WEBDA \\
BH 164 	         &  7.81   & 423.01 	        & 209 	&     2   & 0  &  \cite{2019bossini}\\
BH 221 	         &  8.01   & 1131.22 	        & 162 	&     1   & 0  &    WEBDA \\
BH 23 	         &  7.14   & 443.07 	        & 94 	&     1   & 0  &    WEBDA \\
BH 56 	         &  7.24   & 927.64 	        & 122 	&     3   & 0  &    WEBDA \\
BH 99 	         &  7.908  & 449.44 	        & 389 	&     1   & 0  &  \cite{2019bossini} \\
Biurakan 2 	     &  7.011  & 1845.02 	        & 68 	&     2   & 0  &    WEBDA \\
Blanco 1 	     &  7.975  & 237.53 	        & 381 	&     82  & 12 &  \cite{2019bossini} \\
Bochum 13 	     &  6.823  & 1763.67 	        & 73 	&     1   & 0  &    WEBDA \\
Collinder 106 	 &  6.74   & 1574.80 	        & 114 	&     1   & 0  &    WEBDA \\
Collinder 107 	 &  7.00   & 1620.75 	        & 159 	&     1   & 0  &    WEBDA \\
Collinder 132 	 &  7.08   & 666.22 	        & 99 	&     1   & 0  &    WEBDA \\
Collinder 135 	 &  7.407  & 305.06 	        & 352 	&     1   & 0  &    WEBDA \\
Collinder 197 	 &  7.128  & 967.12 	        & 243 	&     4   & 0  &    WEBDA \\
Collinder 258 	 &  7.834  & 1308.90 	        & 124 	&     1   & 0  &  \cite{2019bossini} \\
DBSB 104 	     &         & 1101.32 	        & 27 	&     1   & 0  &   \\ 
Dolidze 8 	     &         & 1007.05 	        & 43 	&     3   & 0  &   \\ 
FSR 0771 	     & 8.195   & 1618.12 	        & 48 	&     1   & 0  &  \cite{2019bossini} \\
FSR 0951 	     & 8.589   & 1808.32 	        & 195 	&     1   & 0  &  \cite{2019bossini} \\
Gulliver 11 	 &         & 942.51 	        & 64 	&     1   & 0  &  \\
Gulliver 21 	 & 8.472   & 664.89 	        & 126 	&     1   & 0  &  \cite{2019bossini} \\
Gulliver 6 	     &         & 422.48 	        & 343 	&     6   & 2  &   \\
Gulliver 9 	     &  7.154  & 503.78 	        & 265 	&     1   & 0  &  \cite{2019bossini} \\
Haffner 13 	     &  7.497  & 574.05 	        & 210 	&     1   & 0  &  \cite{2019bossini} \\
Harvard 10 	     &  7.899  & 705.22 	        & 164 	&     1   & 0  &  \cite{2019bossini} \\
Harvard 5 	     &  7.812  & 1308.90 	        & 52 	&     1   & 0  &  \cite{2019bossini} \\
IC 1396 	     &  7.054  & 938.09 	        & 460 	&     2   & 0  &    WEBDA \\
IC 2391 	     &  7.561  & 151.93 	        & 224 	&     7   & 0  &  \cite{2019bossini} \\
IC 2395 	     &  7.223  & 725.16 	        & 297 	&     4   & 0  &    WEBDA \\
IC 2488 	     &  8.2    & 1392.76 	        & 435 	&     1   & 0  &  \cite{2019bossini} \\
IC 2602 	     &  7.547  & 152.42 	        & 311 	&     33  & 6  &  \cite{2019bossini}\\
IC 2714 	     &  8.55   & 1390.82 	        & 934 	&     1   & 0  &  \cite{2019bossini} \\
IC 4651 	     &  9.057  & 946.97 	        & 854 	&     8   & 0  &    WEBDA \\
IC 5146 	     &  6.00   & 824.40 	        & 108 	&     2   & 0  &    WEBDA \\
King 6 	         &  8.582  & 742.94 	        & 234 	&     1   & 0  &  \cite{2019bossini}\\
L 1641S 	 &          & 437.06 	        & 101 	&     1   & 0  &   \\
Mamajek 1 	 &  6.778   & 98.72 	        & 20 	&     9   & 5  &    WEBDA \\
Mamajek 4 	 &  8.824   & 449.64 	        & 199 	&     3   & 0  &  \cite{2019bossini} \\
Melotte 111  &   8.652  & 96$^b$                &  859  &   98      & 10    &   WEBDA   \\ 
Melotte 20 	 &  7.938   & 176.43 	        & 764 	&     50  & 0  & \cite{2019bossini} \\
Melotte 22 	 &  7.937   & 136.13 	        & 992 	&     5   & 0  &  \cite{2019bossini} \\
Muzzio 1 	 &          & 1926.78 	        & 40 	&     1   & 0  &  \\
NGC 129 	 &  7.886   & 1956.95 	        & 392 	&     1   & 0  &    WEBDA \\
NGC 1333 	 &          & 299.04 	        & 50 	&     3   & 0  &  \\
NGC 1545 	 &  8.019   & 731.53 	        & 139 	&     1   & 0  & \cite{2019bossini} \\
NGC 1579 	 &          & 552.49 	        & 56 	&     2   & 0  &  \\ 
NGC 1662 	 &   8.957  & 416.67 	        & 238 	&     24  & 2  &  \cite{2019bossini} \\
NGC 188 	 &   9.632  & 1972.39 	        & 864 	&     3   & 0  &    WEBDA \\
NGC 1901 	 &  8.918   & 424.09 	        & 75 	&     15  & 0  &  \cite{2019bossini} \\
NGC 2232 	 &   7.87   & 326.05 	        & 198 	&     2   & 0  &    WEBDA \\
NGC 2244 	 &  6.896   & 1620.75 	        & 623 	&     13  & 0  &    WEBDA \\
NGC 2287 	 &  8.486   & 735.29 	        & 645 	&     1   & 0  &  \cite{2019bossini} \\
NGC 2323 	 &   7.975  & 1003.01 	        & 866 	&     1   & 0  & \cite{2019bossini} \\
NGC 2354 	 &   8.126  & 1328.02 	        & 276 	&     2   & 0  &    WEBDA \\
NGC 2422 	 &  8.104   & 483.09 	        & 442 	&     43  & 1  &  \cite{2019bossini} \\
NGC 2423 	 &   8.991  & 956.94 	        & 430 	&     1   & 0  &  \cite{2019bossini} \\
NGC 2437 	 &   8.390  & 1658.37 	        & 1797 	&     1   & 0  &    WEBDA \\
NGC 2451A 	 &   7.647  & 193.61 	        & 337 	&     26  & 1  &  \cite{2019bossini} \\
NGC 2451B 	 &  7.592   & 367.78 	        & 298 	&     15  & 2  &  \cite{2019bossini} \\
NGC 2516 	 &  8.4     & 413.74 	        & 798 	&     144 & 1  &  \cite{2019bossini} \\
NGC 2546 	 &   7.962  & 967.12 	        & 165 	&     1   & 0  &  \cite{2019bossini} \\
NGC 2547 	 &  7.432   & 391.70 	        & 233 	&     4   & 0  &  \cite{2019bossini} \\
NGC 2548 	 &  8.557   & 775.80 	        & 479 	&     37  & 1   &    WEBDA \\
NGC 3114 	 &  8.093   & 1048.22 	        & 1296 	&     6   & 0  &    WEBDA \\
NGC 3532 	 & 8.601    & 484.03 	        & 1889 	&     11  & 0  &  \cite{2019bossini} \\
NGC 3680 	 &  9.077   & 1071.81 	        & 100 	&     8   & 0  &    WEBDA \\
NGC 5662 	 &  7.968   & 776.40 	        & 255 	&     1   & 0  &    WEBDA \\
NGC 5822 	 &  8.95    & 842.46 	        & 667 	&     1   & 0  &  \cite{2019bossini} \\
NGC 6124 	 &   8.147  & 642.26 	        & 1327 	&     1   & 0  &    WEBDA \\
NGC 6178 	 &  7.248   & 909.92 	        & 45 	&     1   & 0  &    WEBDA \\
NGC 6193 	 &  6.775   & 1231.53 	        & 465 	&     2   & 0  &    WEBDA \\
NGC 6231 	 &  6.843   & 1697.79 	        & 653 	&     11  & 0  &   WEBDA \\
NGC 6242 	 &  7.767   & 1324.50 	        & 523 	&     1   & 0  &  \cite{2019bossini} \\
NGC 6250 	 &  7.415   & 1012.15 	        & 85 	&     1   & 0  &    WEBDA \\
NGC 6281 	 &  8.497   & 533.90 	        & 513 	&     6   & 0  &    WEBDA \\
NGC 6475 	 &  8.477   & 279.96 	        & 941 	&     29  & 0  &  \cite{2019bossini} \\
NGC 6811 	 &   8.936  & 1149.43 	        & 306 	&     4   & 0  &  \cite{2019bossini} \\
NGC 6866 	 &   8.89   & 1457.73 	        & 104 	&     1   & 0  &  \cite{2019bossini} \\
NGC 6871 	 &  6.958   & 1945.53 	        & 594 	&     11  & 0  &    WEBDA \\
NGC 6910 	 &  7.127   & 1834.86 	        & 159 	&     2   & 0  &    WEBDA \\
NGC 6913 	 &  7.111   & 1808.32 	        & 89 	&     2   & 0  &    WEBDA \\
NGC 6940 	 &   9.01   & 1055.97 	        & 593 	&     4   & 0  &  \cite{2019bossini} \\
NGC 7092 	 & 8.491    & 299.67 	        & 161 	&     6   & 0  &  \cite{2019bossini} \\
NGC 7129 	 &          & 920.81 	        & 36 	&     1   & 0  &   \\ 
NGC 7243 	 &  8.006   & 896.06 	        & 313 	&     2   & 0  & \cite{2019bossini} \\
NGC 743 	 &          & 1135.07 	        & 62 	&     1   & 0  &  \\ 
NGC 752 	 &  9.05    & 446.63 	        & 240 	&     9   & 0  &    WEBDA \\
Pismis 5 	 &  7.197   & 974.66 	        & 92 	&     1   & 0  &    WEBDA \\
Platais 3 	 &  8.319   & 177.94 	        & 90 	&     12  & 0  &  \cite{2019bossini} \\
Platais 8 	 &  7.75    & 134.86 	        & 211 	&     13  & 3  &    WEBDA \\
Platais 9 	 &  7.894   & 183.05 	        & 128 	&     18  & 4  &  \cite{2019bossini} \\
Pozzo 1 	 &  7.117   & 350.51 	        & 390 	&     3   & 0  &  \cite{2019bossini} \\
Roslund 3 	 &  7.643   & 1709.40 	        & 112 	&     1   & 0  &  \cite{2019bossini} \\
RSG 5 	     &  7.70    & 339.56 	        & 173 	&     1   & 0  &    \cite{2016roser} \\
RSG 7 	     &  8.30    & 427.90 	        & 133 	&     1   & 0  &    \cite{2016roser} \\
RSG 8 	     &  8.50    & 452.49 	        & 211 	&     1   & 0  &    \cite{2016roser}\\
Ruprecht 91  &  6.90    & 1078.75 	        & 214 	&     2   & 0  &    WEBDA \\
Stephenson 1 &  7.435   & 359.84 	        & 97 	&     3   & 0  & \cite{2019bossini}  \\
Stock 12 	 &  8.188   & 442.28 	        & 109 	&     2   & 0  & \cite{2019bossini}  \\
Stock 1 	 &  8.676   & 408.66 	        & 183 	&     2   & 0  & \cite{2019bossini}  \\
Stock 23 	 &  7.973   & 619.96 	        & 89 	&     3   & 0  & \cite{2019bossini} \\
Stock 2 	 &  8.23    & 378.64 	        & 1190 	&     1   & 0  &     WEBDA \\
Stock 5 	 &  7.73    & 987.17 	        & 115 	&     1   & 0  &     WEBDA \\
Stock 7 	 &  7.214   & 693.96 	        & 141 	&     1   & 0  &     WEBDA\\
Teutsch 38 	 &  8.14    & 665.78 	        & 109 	&     1   & 0  &     WEBDA\\
Trumpler 10  &  7.74    & 437.25 	        & 478 	&     2   & 0  &  \cite{2019bossini} \\
Trumpler 2 	 &  7.962   & 698.81 	        & 179 	&     2   & 0  &  \cite{2019bossini} \\
Trumpler 3 	 &  7.83    & 684.46 	        & 222 	&     2   & 0  &     WEBDA \\
Turner 5 	 &  8.49    & 421.41 	        & 28 	&     1   & 0  &     WEBDA \\
vdBergh 1 	 &  8.025   & 1912.05 	        & 73 	&     1   & 0  &     WEBDA \\
\end{longtable}
\end{center}
\end{document}